\documentclass[fleqn,usenatbib]{mnras}

\usepackage[T1]{fontenc}
\usepackage{ae,aecompl}

\usepackage{graphicx}	
\usepackage{amsmath}	
\usepackage{amssymb}	
\usepackage{bm}
\usepackage{tabularx} 
\usepackage{color}
\usepackage{times}
\usepackage{hyperref}
\usepackage[normalem]{ulem} 

\makeatletter
\newcommand{\HI}{{\rm H\,{\scriptstyle I}}}
\newcommand{\HII}{{\rm H\,{\scriptstyle II}}}

\newcommand{\Rmnum}[1]{\expandafter\@slowromancap\romannumeral #1@}

\newcommand{\xHI}{x_{\mbox{\tiny H\Rmnum{1}}}}
\newcommand{\xHII}{x_{\mbox{\tiny H\Rmnum{2}}}}

\makeatother

\title[Recovering the $\HII$ region size statistics from 21-cm tomography]
  {Recovering the $\HII$ region size statistics from 21-cm tomography}

\author[K. Kakiichi et al.] 
{Koki Kakiichi$^{1,2}$\thanks{E-mail: k.kakiichi@ucl.ac.uk}, 
Suman Majumdar$^3$,
Garrelt Mellema$^4$, 
Benedetta Ciardi$^2$,
\newauthor 
Keri L. Dixon$^5$,
Ilian T. Iliev$^5$,
Vibor Jeli\'{c}$^{7,8}$,
L\'eon V. E. Koopmans$^6$,
Saleem Zaroubi$^6$,
\newauthor
and Philipp Busch$^2$ 
\vspace{0.1in}\\
$^1$ Department of Physics and Astronomy, University College London, London, WC1E 6BT, UK \\
$^2$ Max Planck Institute for Astrophysics, Karl-Schwarzschild Stra\ss e 1, 85741 Garching, Germany \\  
$^3$ Department of Physics, Blackett Laboratory, Imperial College, London SW7 2AZ, UK\\
$^4$ Department of Astronomy and Oskar Klein Centre, AlbaNova, Stockholm University, SE-10691 Stockholm, Sweden\\
$^5$ Astronomy Centre, Department of Physics and Astronomy, Pevensey II Building, University of Sussex, Falmer, Brighton BN1 9QH, UK \\
$^6$ Kapteyn Astronomical Institute, University of Groningen, PO Box 800, 9700AV Groningen, the Netherlands \\
$^7$ Ru{\dj}er Bo\v{s}kovi\'{c} Institute, Bijeni\v{c}ka cesta 54, 10000 Zagreb, Croatia \\
$^8$ ASTRON, The Netherlands Institute for Radio Astronomy, PO Box 2, 7990 AA Dwingeloo, the Netherlands
}
\begin{document}
\label{firstpage}
\pagerange{\pageref{firstpage}--\pageref{lastpage}} \pubyear{2016}
\maketitle


\begin{abstract}
We introduce a novel technique, called ``granulometry", to characterize and recover the mean size and the size distribution of $\HII$ regions from 21-cm tomography. The technique is easy to implement, but places the previously not very well defined concept of morphology on a firm mathematical foundation. The size distribution of the cold spots in 21-cm tomography can be used as a direct tracer of the underlying probability distribution of $\HII$ region sizes. We explore the capability of the method using large-scale reionization simulations and mock observational data cubes while considering capabilities of SKA1-low and a future extension to SKA2. We show that the technique allows the recovery of the $\HII$ region size distribution with a moderate signal-to-noise ratio from wide-field imaging ($\rm SNR\lesssim3$), for which the statistical uncertainty is sample variance dominated. We address the observational requirements on the angular resolution, the field-of-view, and the thermal noise limit for a successful measurement. To achieve a full scientific return from 21-cm tomography and to exploit a synergy with 21-cm power spectra, we suggest an observing strategy using wide-field imaging (several tens of square degrees) by an interferometric mosaicking/multi-beam observation with additional intermediate baselines ($\sim2-4\rm~km$) in a SKA phase 2.
\end{abstract}


\section{Introduction}
The 21-cm signal arising from the spin flip transition in the neutral
hydrogen ($\HI$) atoms promises to unearth an unprecedented amount of
information about the Cosmic Dawn (CD) and the Epoch of Reionization
(EoR). It will help us to resolve many outstanding questions regarding
this particular phase of the history of our universe such as: How
did reionization progress with time? What was the morphology of $\HI$
distribution in the intergalactic medium (IGM) during the different
phases of reionization? Which were the major sources that drive
reionization? (for reviews see,
e.g. \citealt{2006PhR...433..181F}, \citealt{2010ARA&A..48..127M} and \citealt{2012RPPh...75h6901P}). 

The
first generation of radio interferometers such as the
GMRT\footnote{http://gmrt.ncra.tifr.res.in/} \citep{paciga13},
LOFAR\footnote{http://www.lofar.org} \citep{yatawatta13},
MWA\footnote{http://www.mwatelescope.org} \citep{bowman13} and
PAPER\footnote{http://eor.berkeley.edu} \citep{ali15} target the
detection of this signal through statistical estimators such as the
power spectrum and variance. While these statistical estimators
ensure an optimum signal-to-noise ratio (SNR) to increase the detection
possibility, it will not be possible to characterize the
signal completely through these measurements alone, as the signal from
the EoR is expected to be highly non-Gaussian in nature (e.g.
\citealt{2003ApJ...596....1C,2004ApJ...613...16F,bharadwaj05a,2005MNRAS.363.1049C,2006MNRAS.369.1625I,2006MNRAS.372..679M,2009MNRAS.393.1449H,2015MNRAS.449L..41M,2016MNRAS.456.3011D}). As a result, many completely different underlying $\HI$ distributions could give rise to the same power spectrum and variance for the signal, which would lead to a large degeneracy and ambiguity in the estimated astrophysical parameters through these statistics. This is why it is essential to quantify and characterize the signal beyond these one- and two-point statistics. 

One way to achieve this is through a hierarchy of the higher order correlation functions (such as three-point and four-point correlation functions) and their corresponding Fourier analogues \citep{bharadwaj05a,2008PhRvD..77j3506C,2007ApJ...662....1P,2015PhRvD..92h3508M,2015MNRAS.449L..41M,2017MNRAS.464.2992M}, which can be directly estimated using the observed visibilities in a radio interferometer.

Another alternative to understanding and characterizing the signal more fully is through the images of the 21-cm signal (thus the $\HI$ distribution in the IGM) at different stages of reionization. This is more popularly known as `21-cm Tomography' (see e.g. \citealt{1997ApJ...475..429M}), and it 
will essentially provide us with a time lapse movie of the state of the $\HI$ in a certain portion of the sky during the EoR.  As radio interferometry is only sensitive to the fluctuations in the signal, not its absolute value,  regions of ionized hydrogen will appear as negative `cold spots' in these images. Thus, studying the properties of cold spots in the 21-cm tomographic data will provide a direct insight into ``the morphology of reionization'', which is characterized by the shapes, sizes, and spatial distribution of the $\HII$ regions. These studies will help us to directly quantify the ionized volume-filling fraction, the mean size, and the bubble size distribution of the $\HII$ regions. 

For this reason, the 21-cm tomographic imaging during the CD and the EoR has been identified as one of the major science goals of the future radio interferometer Square Kilometre Array\footnote{https://www.skatelescope.org/} (SKA) \citep{2015aska.confE...1K,2015aska.confE..10M,2015aska.confE..15W}.
Even currently operating interferometers such as LOFAR and MWA are expected to produce relatively coarser resolution images at a moderate SNR \citep{2012MNRAS.425.2964Z,2013ApJ...767...68M}. However, the tomographic data cubes still need to be analysed in a statistical sense in order to derive astrophysical parameters.

The data analysis strategy of 21-cm tomographic images is a relatively unexplored field. Previous works have focused on topological analysis of 21-cm tomography through the Minkowski functionals \citep{2006MNRAS.370.1329G,2008ApJ...675....8L,friedrich11,2014JKAS...47...49H,2016arXiv160202351Y} or by extracting the size, shape, and location of the $\HII$ region using the matched filter technique on the raw observed visibilities and images   \citep{datta07,datta08,datta09,datta12,datta16,majumdar11,majumdar12,2013ApJ...767...68M}. The latter have explored both targeted (where a very bright high redshift ionizing source has already been identified through observations at other wavelengths) and blind (where the location of the ionizing source is not known) searches for $\HII$ regions on the observed visiblities or on the image planes. For example, \citet{2013ApJ...767...68M} have demonstrated that by using this matched filtering method it is also possible to extract the size distribution of the 21-cm cold spots from an observational data set. Some additional work has been done on studying size distributions in simulations of reionizations \citep{2011MNRAS.413.1353F, 2016MNRAS.461.3361L}, but the application of these techniques to 21-cm tomography has not yet been explored.

One of the main challenges in recovering the statistical measures of the morphology (e.g. shape or size distribution) of the $\HII$
regions from 21-cm tomography is due to the fact that there is no unique definition of the 
`size or shape of an $\HII$ region'. Several methods and definitions of $\HII$ region size and shape have been 
used to extract the size distribution form simulated 21-cm data (\citealt{2006MNRAS.369.1625I,2007ApJ...654...12Z,2007MNRAS.377.1043M,2007ApJ...669..663M,friedrich11,majumdar14,2016MNRAS.461.3361L}).  So far, there does not seem to be a consensus on how these methods can actually trace the underlying size distribution of the $\HII$ regions. Apart from the ambiguity of the definition of an $\HII$ region, the other major challenge in obtaining the size distribution is to identify the `true' ionized cells from all the cold spots present in the 21-cm tomographic data. As radio interferometers can only capture fluctuations in the signal, the tomographic data will have mean subtracted signal represented by positive and negative pixels in the image plane at a certain redshift. A pixel containing neutral gas but having density below the mean density could appear as a negative pixel. In addition, the mean of the 21-cm signal (which corresponds to the `zero' of the signal in that data volume) will also vary with redshift as reionization progresses from its early to late stages. These two intrinsic features of the observed signal will produce additional confusion in identifying true ionized pixels from 21-cm images. 

The main goal of this paper is to introduce a novel granulometric analysis method in the context of the EoR 21-cm tomography for the first time. We focus on measuring the $\HII$ region size distribution from 3D data cube and 2D images, hence, directly characterizing the aspect of the morphology of $\HII$ regions. This granulometric technique is based on a well-defined method in mathematical morphology and stochastic geometry (\citealt{Serra1983,Dougherty,Chiu2013}), which provides a well-formulated definition and the characterization of `size' of the objects based on the idea of sieving (\citealt{Matheron}). We further investigate the possibility and requirement to recover the $\HII$ region size distribution in 21-cm tomography from future radio interferometric observations.

This paper is organized as follows. Section~\ref{sec:models} describes the 21-cm and reionization models. Section~\ref{sec:data_analysis_method} introduces granulometry in the context of our data analysis methodology. Section~\ref{sec:mock_obs} describes our mock data cube under a simple model for radio interferometry. Section~\ref{sec:results_granulometry} presents the theoretical aspects of the granulometric analysis of 21-cm tomography. Section~\ref{sec:result_3d_granulometry} and \ref{sec:result_2d_granulometry} presents the observational prospects and requirements for measuring the $\HII$ region size distribution using 3D image cubes and 2D image slices from 21-cm tomography. The implications for baseline design and observing strategy are discussed in Section~\ref{sec:discussions}. Finally, our conclusions are summarized in Section~\ref{sec:conclusions}. 

\section{EoR 21-cm signal}\label{sec:models}

The spin flip transition of the hydrogen ground state corresponds to the 21-cm line with $\nu_{21}=1420.4\rm~MHz$, whose emission or
absorption at redshift $z$ is observed at a frequency
$1420.4(1+z)^{-1}\rm MHz$. The differential brightness temperature of
21-cm line against the Cosmic Microwave Background (CMB) is given by
(e.g. \citealt{1958PIRE...46..240F,1959ApJ...129..536F,1997ApJ...475..429M,2012RPPh...75h6901P})
\begin{equation}
\delta T_{21}=T_0(z)(1+\delta_{\rm b})\xHI
\left(1-\frac{T_{\rm CMB}(z)}{T_{\rm S}}\right),
\label{eq:dTb_definition}
\end{equation}
where the pre-factor $T_0(z)\approx27{\rm~mK}\left(\frac{\Omega_{\rm b} h^2}{0.023}\right)\left(\frac{0.15}{\Omega_{\rm m}h^2}\frac{1+z}{10}\right)^{1/2}$ depends on cosmological parameters and redshift. For simplicity, we ignore the effect of redshift-space distortions due to the peculiar velocities of matter along the line of sight. 

The spin temperature $T_{\rm S}$ characterizes the relative populations of the two spin states. This population is determined by collisions and radiative excitations. Here we assume that the spin temperature is coupled to the kinetic temperature of gas and that the gas temperature is always above the CMB temperature. These conditions are expected to be valid for most of the reionization period (likely $z<12$). This simplifies Equation~(\ref{eq:dTb_definition}) to
\begin{equation}
\delta T_{21}=T_0(z)(1+\delta_{\rm b})\xHI,\label{eq:21}
\end{equation}
which only depends on the neutral fraction of hydrogen and the baryon density fluctuations. We adopt this so-called full coupling approximation throughout the paper.

\subsection{Models for 21-cm maps from EoR}
To demonstrate the use of granulometric analysis in 21-cm tomography, we employ two models to characterize the distribution of ionized regions: (1) a toy model in which we impose a log-normal bubble size distribution and (2) the results from detailed radiative transfer (RT) simulations. 

\subsubsection{Log-normal bubble model}\label{lognormalbubbles}
In this simple approach, the reionization process is modelled as a percolation of spherical $\HII$ regions in a homogeneous IGM density field. This simple model is constructed by randomly distributing spherical ionized bubbles in a simulation box, and does not include any correlation of the bubble distribution with the underlying density field. Motivated by \citet{2007MNRAS.377.1043M} and \citet{2011MNRAS.413.1353F}, the sizes of the radii of $N_b$ bubbles are randomly drawn from a log-normal distribution of the bubble sizes $R$,
\begin{equation}
\frac{dP(R)}{dR}=\frac{1}{\sqrt{2\pi\sigma_{\rm R}^2}R}\exp\left[-\frac{(\ln R-\ln \bar{R})^2}{2\sigma_{\rm R}^2}\right],
\label{lognormal}
\end{equation}
where $N_b$, $\bar{R}$ and $\sigma_{\rm R}$ are the three parameters of the model. The ionization profile of each bubble is assumed to be a spherical top-hat function $\xHII(\mathbf{r})=\Pi(\mathbf{r}-\mathbf{r}_i|R_{i})$ of radius $R_i$ centred at a position $\mathbf{r}_i$. Note that when multiple bubbles overlap we take the maximum of all the ionized fractions ($=1$).

Finally we use Monte Carlo realizations of the log-normal bubble model in a volume $V_{\rm box}=(1h^{-1}\rm cGpc)^3$ with $256^3$ cells and $3.9h^{-1}\rm cMpc$ resolution to create a differential brightness temperature map.

\subsubsection{Radiative transfer simulation}\label{sec:simulation}

To demonstrate our new data analysis methodology for 21-cm tomography we use snapshots from a full RT simulation performed within the PRACE4LOFAR project\footnote{This project was executed using two allocations
of Tier-0 time, 2012061089 and 2014102339, awarded by the Partnership for Advanced Computing in Europe (PRACE).}. Reionization is modelled by post-processing an $N$-body simulation with a RT calculation. 
The $N$-body code used is \textsc{\small CubeP$^3$M} (\citealt{2013MNRAS.436..540H}) and the RT code is \textsc{\small C$^2$-RAY} (\citealt{2006NewA...11..374M}). We assume a flat $\Lambda$CDM cosmology with the WMAP5 parameters $h=0.7,~\Omega_{\rm m}=0.27,~\Omega_\Lambda=0.73,~\Omega_{\rm b}=0.044,~\sigma_8=0.8,~n_{\rm s}=0.96$ (\citealt{2009ApJS..180..330K}), which are also consistent with Planck data (\citealt{2015arXiv150201589P}). 

The simulation we use throughout this paper is similar to the LB1 model in \cite{2016MNRAS.456.3011D}, where details regarding simulation techniques and methods can be found. However, the volume
considered here is a larger $500h^{-1}\rm cMpc$ on each side for which
the $N$-body simulation was run with $6192^3$ particles with mass resolution $4.05\times10^7\rm~M_\odot$ (Dixon et al. in prep). The minimum dark matter halo mass used in the RT
simulation is $1\times10^9\rm~M_\odot$ (25 particles). The RT
simulation was performed on smoothed and gridded density fields consisting of $300^3$
cells. The ionizing photons from the sources (galaxies) are
assumed to linearly scale with the host halo mass such that the number of ionizing photons released into the IGM is $g_\gamma=1.7$ per baryon per $10^7$~years. This RT simulation of reionization runs from $z=21$ to $z=6$.

We use the resulting snapshots of the $\HII$ fraction and the gas density (on a $300^3$ grid) for the analysis presented in this paper. We select the $z=6.8$ snapshot as our fiducial reference redshift unless otherwise stated. At this redshift, the volume-averaged ionized fraction corresponds to $\langle\xHII\rangle_V=0.40$ (see Section~\ref{sec:void_bias} for a discussion on the dependence of our results on  this choice).

\section{Data Analysis Method}\label{sec:data_analysis_method}

This section describes basic concepts involved in our data analysis methodology of 21-cm tomography using the granulometric technique. The same analysis method is used both for noiseless and noisy data.

\subsection{Granulometry}\label{sec:method_granulometry}
Granulometry is a technique in mathematical morphology and image analysis that measures a size distribution of objects (\citealt{Serra1983,Dougherty}). The central idea is based on the concept of sieving (\citealt{Matheron}). This provides a mathematically well-defined measure of the size distribution of a collection of objects, in our case $\HII$ regions and 21-cm cold spots, the latter defined as regions where the differential brightness temperature is less than a certain threshold value. 
The tool we used for our granulometric analysis is publicly available online\footnote{\url{http://www.star.ucl.ac.uk/~kakiichi/codetools.html}}. 

\subsubsection{Binary images and data cubes}

Since we need to define objects, the first step is the creation of a binary field of the quantity of interest (either ionization fraction or 21-cm signal). A binary field of the $\HII$ fraction map, denoted by $X_{\rm HII}(\mathbf{r})$, is defined as:
\begin{align}
X_{\rm HII}(\mathbf{r})= \left\{
  \begin{array}{cc}
    1 & \mbox{if $x_{\rm HII}(\mathbf{r})\geq x_{\rm HII}^\mathrm{th}$},\\
    0 & \mbox{if $x_{\rm HII}(\mathbf{r})<x_{\rm HII}^\mathrm{th}$}.
  \end{array}
  \right.
\end{align}
where $x_{\rm HII}^\mathrm{th}$ is a given ionization threshold.
The ionized regions are marked as ones in the binary field. Throughout this paper we take $x_{\rm HII}^\mathrm{th}=0.5$ to mark the 50 per cent transition between neutral and ionized phase.
 
Similarly, binary fields of the 21 cm signal, denoted by $X_{21}(\mathbf{r})$, are produced introducing a threshold value for the pixels of the mean subtracted 21-cm signals,
\begin{align}
X_{21}(\mathbf{r})= \left\{
  \begin{array}{cc}
    1 & \mbox{if $\Delta T_{21}(\mathbf{r})<\Delta T_{21}^\mathrm{th}$},\\
    0 & \mbox{if $\Delta T_{21}(\mathbf{r})\geq\Delta T_{21}^\mathrm{th}$},
  \end{array}
  \right.
\end{align}
where $\Delta T_{21}=\delta T_{21}-\langle\delta T_{21}\rangle$ is the mean subtracted 21-cm brightness temperature fluctuation. Because we are interested in the size distribution of 21-cm cold spots, we define the pixels below the threshold as ones. Throughout this paper we take $\Delta T_{21}^\mathrm{th}=0$ and we call ``cold spots'' all regions below this value. We discuss the possibility of varying this threshold value in Section \ref{sec:void_bias}. We also define voids in the density field in the same way as 21-cm cold spots, i.e. a connected pixels below the mean, but using the density field instead of 21-cm signal. Note that $\mathbf{r}$ is the Cartesian coordinate converted from the angular and frequency separations in the data cube. 

The total volume filled by 21-cm cold spots or $\HII$ regions is then given by (here the subscript indicating the physical quantity is dropped), 
\begin{equation}
V[X]=\int X(\mathbf{r})d^3r.
\end{equation}
The volume-filling factors of $\HII$ regions and 21-cm cold spots in a 3D tomographic data cube are then given by $Q_{\rm HII}=V[X_{\rm HII}]/V_{\rm box}$ and $Q_{21}=V[X_{21}]/V_{\rm box}$, respectively. The filling factor $Q_{\rm 21}^{\rm 2d}$ for a 2D 21-cm image can be defined in an analogous way.

\subsubsection{Size distribution measurement}
The granulometric analysis technique is applied on the binary field. The basic idea is to probe the field with a so-called structuring element of a certain shape. This (loosely speaking) represents the shape of holes in a sieve. For our analysis, we choose the structuring element to be a sphere of radius $R$ (or an disc of radius $R$ when analysing a 2D image), defined as $S_R=\Pi(\mathbf{r}-\mathbf{r}_0|R)$ where $\mathbf{r}_0$ is the coordinate of the centre of the structuring element. The mathematical formulation of the concept of sieving corresponds to a {\it morphological opening} operation\footnote{This is a well-defined operation in mathematical morphology, which in turn can be formulated in terms of more fundamental set-theoretical operations, Minkowski addition ($\oplus$) and subtraction ($\ominus$), as $X\circ S_R\equiv(X\ominus S_R)\oplus S_R$ for a symmetric structuring element (e.g. \citealt{Dougherty}). For the detailed mathematical foundation, see \cite{Matheron}, \cite{Serra1983}, \cite{Dougherty} and \cite{Chiu2013}. In practice, this opening operation ($\circ$) is easy to use as it is implemented as a part of widely used high-level programming languages and standard libraries such as $\rm python$ and $\rm scipy$ package.} denoted by a symbol $\circ$ (e.g. \citealt{Dougherty}). Hereafter, we refer to the morphological opening operation as sieving. A brief introduction is presented in Appendix~\ref{sec:App}.

The sieving of the binary field $X$ through a `hole' of radius $R$ is thus mathematically expressed as
\begin{equation}
X'(\mathbf{r})=X\circ S_{\rm R}.
\end{equation}
The new binary field $X'(\mathbf{r})$ represents the parts of the original binary field $X$ that {\it remains} after sieving (see Section~\ref{sec:proof} for an example). Structures smaller than the radius of the structuring element are removed from the sieved binary field $X'(\mathbf{r})$. In astrophysical terms, the sieved ionization fraction field or 21-cm image only contains the $\HII$ regions or cold spots larger than the radius of the structuring element.

This concept of sieving defines the size distributions of 21-cm cold spots and $\HII$ regions in a mathematically well-formulated way. In the granulometric analysis the cumulative size distribution $F(<R)$, i.e. the fraction of structures whose size is smaller than a radius $R$, is given by the fraction of volume removed by sieving (e.g. \citealt{Serra1983,Dougherty}), 
\begin{equation}
F(<R)=1-\frac{V[X\circ S_R]}{V[X]}.
\end{equation}
Granulometry is basically a method to measure the size distribution by counting, using successive sieving operations on a binary field with an increasing size of the structuring element. In other words, granulometry counts the number of objects that `fit' in the structure of interest. For example, a large, irregularly connected $\HII$ region is decomposed into a sum of smaller spherical objects, whereas a large spherical $\HII$ region is counted as one object. The number of spherical objects should also be minimized to describe the original structure.

The (differential) size distribution, $dF(<R)/dR$, is given by differentiating the cumulative size distribution. For practical applications, we use the above granulometric analysis on a discrete (pixelized) field. A structuring element then has a discrete radius $R_i=i\times\Delta R$, where $\Delta R=L/N_{\rm pix}$, $L$ is the comoving size of the data cube or image, $N_{\rm pix}$ is the number of pixels per dimension, and $i=1,2,\dots,N_{\rm pix}/2$. The differential size distribution is estimated from the discrete cumulative size distribution,
\begin{equation}
\frac{dF(<R_i)}{dR}\equiv \frac{F(<R_{i+1})-F(<R_i)}{\Delta R}.
\end{equation}
Hereafter, $dF_{21}(<R)/dR$ and $dF_{\rm HII}(<R)/dR$ denote the size distributions of 21-cm cold spots and $\HII$ regions measured from the corresponding binary fields, $X_{21}$ and $X_{\rm HII}$, respectively.

An advantage of the granulometric measure of the size distribution is that it attempts to recover the true underlying probability distribution function of sizes \citep{Serra1983}. In Section~\ref{sec:proof} we verify this property using an explicit example. In the terminology  of \citet{2016MNRAS.461.3361L}, the granulometry method is an unbiased estimator of the true size distribution\footnote{It is obvious that for the cases of non-overlapping spherical or circular regions, the granulometric measure returns the true size distribution.}. We compare the granulometric method with other size estimators in Section~\ref{sec:comparison}.

\subsubsection{Moments of size distributions and the normalization}

The mean size of $\HII$ regions or cold spots is given as the first moment of the size distribution,
\begin{equation}
\langle R\rangle=\int_0^\infty R\frac{dF(<R)}{dR}dR.
\end{equation}
The higher-order moments of a size distribution can be defined in a similar way to characterize the variance and skewness of the distribution.

In addition, as we will show in Section~\ref{sec:results_granulometry}, it is convenient to normalize the size distribution with the volume-filling factor such that 
\begin{equation}
\frac{dQ(<R)}{dR}\equiv Q\frac{dF(<R)}{dR}.
\end{equation}
When the size distribution is normalized to the volume-filling factor, one can interpret this quantity as the fraction of the total volume-filling factor contributed by regions of size $R$.

\subsection{Error estimation}\label{sec:error}
So far, we only introduced the theory for granulometric analysis. When measuring the $\HII$ region size distributions from observation of 21-cm cold spots, we must also assess the associated error as any measurement comes with uncertainty. The statistical uncertainties of the results come both from the sample variance of the signal and the thermal noise of the instrument. We estimate the error in the measured size distribution of 21-cm cold spots using an ensemble of mock data cubes with noise.

The thermal noise covariance matrix for the cold-spot size distribution is calculated using $N_{\rm noise}=100$ Monte Carlo realizations of noise (Section~\ref{sec:noise}). Each independent noise cube is added to a 21-cm data cube. We then perform the measurement for each of these noise-added mock data cubes. The noise covariance matrix $C_{ij}^N$ for each pair of pixelized radius bins, $R_i$ and $R_j$, is given by
\begin{align}
C_{ij}^N=\frac{1}{N_{\rm noise}}\sum_{n=1}^{N_{\rm noise}} \left[\left.\frac{dQ(<R_i)}{dR}\right|_n-\left\langle\frac{dQ(<R_i)}{dR}\right\rangle\right]\times \nonumber \\
~~~\left[\left.\frac{dQ(<R_j)}{dR}\right|_n-\left\langle\frac{dQ(<R_j)}{dR}\right\rangle\right],
\end{align}
where $\left\langle dQ(<R_i)/dR\right\rangle$ is the average over $N_{\rm noise}$ realizations.

The sample variance of the error covariance matrix is calculated using the sub-volume method (\citealt{2009MNRAS.396...19N}). For the analysis of a full 3D data cube, a mock 21-cm data cube is split into $N_{\rm sample}=8$ equal sub-volumes. We calculate the covariance matrix of each sub-volume, and the total covariance matrix of the full volume is estimated as an average of the matrices after re-scaling by $N^{-1/2}_{\rm sample}$. For the analysis of a 2D image, we randomly select $N_{\rm sample}=100$ slices along the random principal axis. We repeatedly perform the measurement using the random sub-samples. The sample variance of the covariance matrix $C_{ij}^S$ for each pair of pixelized radius bins, $R_i$ and $R_j$, is given by
\begin{align}
C_{ij}^S=\frac{1}{N_{\rm sample}}\sum_{m=1}^{N_{\rm sample}} \left[\left.\frac{dQ(<R_i)}{dR}\right|_m-\left\langle\frac{dQ(<R_i)}{dR}\right\rangle\right]\times \nonumber \\
\left[\left.\frac{dQ(<R_j)}{dR}\right|_m-\left\langle\frac{dQ(<R_j)}{dR}\right\rangle\right].
\end{align}

The total error covariance matrix is then given by $C_{ij}=C_{ij}^N+C_{ij}^S$. The error is estimated according to the covariance matrix. Note that the sample variance error, which scales with the survey volume as $\propto V_{\rm survey}^{-1/2}$, typically dominates. The thermal noise error has a more complicated dependency on the survey volume as it couples with the sample variance error (see Section~\ref{sec:coupling}), but it is usually a small contribution to the total error budget. We present $1\sigma$ error bars in this paper unless otherwise stated.

\section{Mock interferometric observations}\label{sec:mock_obs}

In this section we describe how we construct our mock data cubes, which represent the signal as observed with an idealised SKA-like radio interferometer. 

\subsection{Instrument}\label{sec:instrument}
We consider EoR experiments with an SKA-like radio interferometer. Our reference array configuration is based on the SKA1-low baseline design documented by Dewdney (2016)\footnote{SKA-TEL-SKO-0000002 (revision 3) accessed on 28th November 2016. \url{http://astronomers.skatelescope.org/wp-content/uploads/2016/05/SKA-TEL-SKO-0000002_03_SKA1SystemBaselineDesignV2.pdf}}. SKA1-low has a frequency coverage between $50\rm~MHz$ and $350\rm~MHz$, corresponding to the redshift range $3\lesssim z\lesssim27$. The assumed design consists of a total of 512 stations. The final configuration is still being developed but here we assume that approximately half of the stations are distributed within a radius of $\sim1\rm~km$; therefore, we assume that the number of core stations is  $N_{\rm st}=256$. We assume a complete uv coverage within a baseline length of $2\rm~km$ (we refer to this as the maximum baseline length $b_{\max}$). The station diameter is taken to be $D_{\rm st}=35\rm~m$. We assume an idealised instrument with an effective area which below the critical frequency $\nu_\mathrm{crit}$ is the same as the geometric area of the station $A_{\rm eff}=\pi(D_{\rm st}/2)^2$, and above it falls off as $(\nu_\mathrm{crit}/\nu)^2$ where $\nu_{\rm crit}=110~\rm MHz$. The system temperature is given by the sum of the receiver noise, $T_{\rm rcvr}=40\rm~K$, and the sky temperature $T_{\rm sky}=60(\nu/300{\rm~MHz})^{-2.55}{\rm~K}$. We refer to this idealised instrument as `SKA1-low'. The instrumental parameters are summarized in Table~\ref{table:instrument}.

In addition, we consider a future SKA2-like instrument extending the core array of SKA1-low to a 2 km radius, and achieving the complete uv-coverage within the maximum baseline of $b_{\rm max}=4\rm~km$. This improves the angular resolution by a factor of 2, but keeps the field-of-view of the single pointing the same as the SKA1-low.  We refer to this instrument as `SKA2'\footnote{Note however that the design of the real phase 2 of SKA has not been determined yet and may not involve an increase of the core size. An alternative example could be an increase of the sensitivity for the same core size as SKA1-low.}.

\begin{table}
\centering
\caption{SKA1-low (and SKA2) instrument parameters. } \label{table:instrument}
    \begin{tabular}{ll}
    \hline
    Configuration				& 				\\
    \hline
    Frequency coverage	& 50 MHz -- 350 MHz	\\
    Number of core stations  $N_{\rm st}$		& 256		\\
    Max. baseline $b_{\rm max}$				& 2 km$^{\dagger}$	\\
    Station diameter $D_{\rm st}$ 	& 35 m			\\
    Effective area $A_{\rm eff}$          & 962 m$^2$ \\
    Critical frequency $\nu_\mathrm{crit}$ & 110 MHz\\
    System temperature $T_{\rm sys}$      & $40{\rm~K}+60(\nu/300{\rm~MHz})^{-2.55}{\rm~K}$ \\
    Angular resolution $\theta_{\rm A}$		& $2.58(\nu/200\rm{~MHz})^{-1}$armin \\
    Primary beam's FWHM			& $3.12(\nu/200\rm{~MHz})^{-1}$deg \\
    \hline
    \multicolumn{2}{m{7cm}}{$^{\dagger}$ Our model SKA2 increases the maximum baseline to $4\rm~km$.}

    \end{tabular}
\end{table}

\subsection{21-cm signal}\label{sec:21-cm_mock}

The 21-cm signal is calculated following equation (\ref{eq:21}) using the results of the RT simulations described in Section~\ref{sec:simulation}. Our fiducial analysis uses the redshift snapshot at $z=6.8$ (corresponding to a frequency of $182\rm~MHz$), where the simulation box has an extension of $\sim4.6\rm~deg$ on a side at this redshift. The spatial resolution is $\Delta r=1.67h^{-1}\rm cMpc$. Each pixel therefore corresponds to an angular size of $0.92\rm~arcmin$. We construct a simulated 21-cm data cube using a coeval snapshot from the RT simulation\footnote{For simplicity, we did not use a lightcone in this work. Although lightcone cubes including redshift-space distortions must be used for a more sophisticated assessment, we wanted to avoid extra complications because our main point is to introduce the new granulometric analysis techniques in 21-cm tomography. The size of the coeval volume corresponds to 43.6 MHz in the frequency direction, which corresponds to a light cone extending from $z=6.1$ to $z=8.13$ if centred at $z=7$. We are effectively assuming that the evolution of the $\HI$ structure is slow enough during this period so that we can use the entire frequency range to approximately represent the state of $z=7$.}. Each 2D slice is  one pixel width and it corresponds to a bandwidth $B=\nu_{21}H(z)\Delta r/[c(1+z)^2]\approx0.15\rm~MHz$, where $H(z)$ is the Hubble parameter and $c$ is the speed of light. 

\subsection{Angular and frequency resolution}\label{sec:uv_mask}

The angular resolution of the radio interferometric observation is characterized by the maximum baseline of the array as $\theta_{\rm A}=\lambda/b_{\rm max}$~radians (see Table~\ref{table:instrument}). For SKA2 the angular resolution is increased by a factor of 2. We implement the angular response of the interferometer by convolving with a Gaussian point spread function (PSF), $R(\theta)$, with FWHM corresponding to $\theta_{\rm A}=2.58(\nu/200\rm{~MHz})^{-1}\rm armin$.

The frequency resolution is determined by the design of the instrument, and for SKA1-low it is expected to be better than 1~kHz. However, in practice, when analysing the signal a lower frequency resolution is used to increase the SNR. Here we assume that the data is smoothed in the frequency direction with a Gaussian kernel of exactly the same physical size as the angular PSF. For the chosen redshift this implies a FWHM of 453~kHz.
 
\subsection{Noise}\label{sec:noise}

The point-source sensitivity of an interferometer is given by (e.g. \citealt{2001isra.book.....T}, eq.~(6.62))
\begin{equation}
\sigma_{\rm S}=\frac{2k_{\rm B}T_{\rm sys}}{\epsilon A_{\rm eff}\sqrt{N_{\rm st}(N_{\rm st}-1)Bt_{\rm int}}},
\end{equation}
where $t_{\rm int}$ is the integration time of an observation and $\epsilon$ is the efficiency factor described in Section~\ref{sec:instrument}. For imaging (i.e. 21-cm tomography), we are concerned with the rms brightness temperature sensitivity (in units of $\rm K$) of an image at angular scale $\Omega_{\rm A}=(\pi/4\ln2)\theta_{\rm A}^2$  (\citealt{Condon}),
\begin{align}
&\sigma_{\rm N}=\left(\frac{\sigma_{\rm S}}{\Omega_{\rm A}}\right)\frac{\lambda^2}{2k_{\rm B}} \label{eq:noise}\\
&~~~\approx7.85\left(\frac{t_{\rm int}}{1000{\rm~hr}}\right)^{-1/2}\left(\frac{B}{0.453{\rm~MHz}}\right)^{-1/2}\left(\frac{\theta_{\rm A}}{2.83'}\right)^{-2}{\rm mK}, \nonumber
\end{align}
at the observed frequency $182\rm~MHz$ ($z=6.8$) and on the scale of the maximum angular resolution element of the SKA1-low\footnote{We are aware that this estimate of the SKA1-low sensitivity might be optimistic. When a more realistic set-up of interferometric imaging is taken into account, to achieve $\sim3-5\rm~mK$ rms noise level it could take an integration time longer than what estimated here. The most important parameter that directly affects our analysis and conclusion is the rms sensitivity, $\sigma_{\rm N}$. The integration time must be regarded only as a rule-of-thumb. Therefore, we quote the rms sensitivity rather than the integration time in this paper.}. This noise estimate is somewhat optimistic as it assumes full uv coverage and more detailed calculations suggest noise levels which are between a factor 1.5 and 2 higher.

As explained in Section~\ref{sec:error}, we produce 100 Monte Carlo realizations of the noise cubes. We generate white (Gaussian) noise fields, which have the same spatial (angular) scale and frequency range as the 21-cm data cube (Section~\ref{sec:21-cm_mock}). The rms noise level $\langle(\Delta T_{\rm N})^2\rangle$ at the scale of the resolution element is then normalized according to equation (\ref{eq:noise}). Because of the assumption of white noise, the noise power spectrum scales as $\Delta_{\rm N}^2(k)\propto k^3$.

To quantify the image quality, we define the SNR for a data cube as the ratio of the rms fluctuations between an image cube and a noise cube on the scale of resolution element ${\rm SNR}(\theta_{\rm A})=\sqrt{\langle (\Delta T_{21})^2\rangle/\langle (\Delta T_{N})^2\rangle}$. 

\subsection{Foregrounds}

We assume that the various foreground signals are perfectly removed from our data cube. \cite{2015aska.confE...5C} discussed the effect of different foreground removal techniques on the reconstructed 21-cm images, showing that good quality reconstructed 21-cm data cubes  are  in principle obtainable. Studying the impact of foreground residuals on the 21-cm tomographic analysis is beyond the scope of this paper. 

\section{Granulometric Analysis}\label{sec:results_granulometry}

In this section, we first present the results of granulometric analysis of one constructed and one simulated distribution of $\HII$ regions, as well as of the noiseless 21-cm signals associated with the latter. The goal is to understand the physical properties probed by the granulometric analysis and how well the 21-cm cold-spot size distribution traces the underlying size distribution of the $\HII$ regions. 

\subsection{A proof-of-concept: log-normal bubble model}\label{sec:proof}

\begin{figure}
  \centering
  \includegraphics[angle=0,width=\columnwidth]{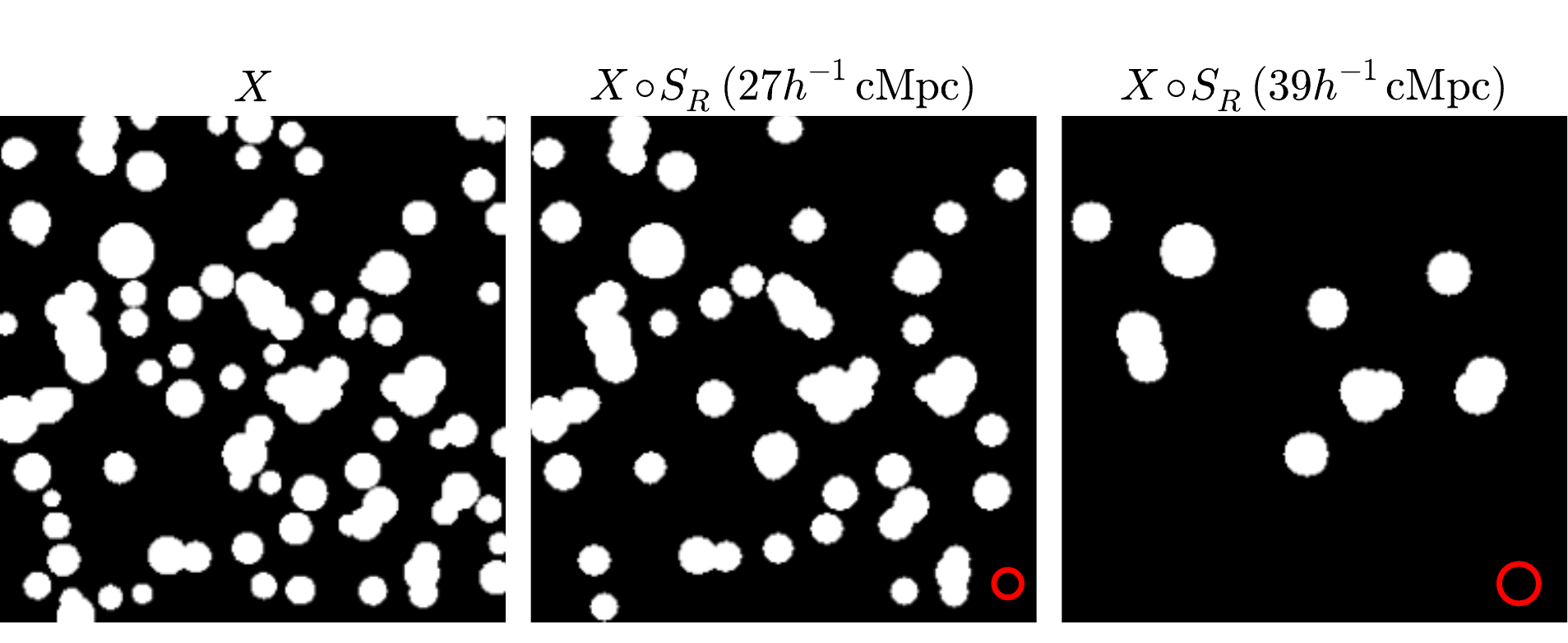}
  \caption{Example of sieving for a 2D log-normal bubble model in a $1h^{-1}\rm cGpc$ box  on a side. The red circle shows the radius of the structuring element. The left panel shows the original (unsieved) distribution of $\HII$ regions. The middle and right panels show the distributions obtained by sieving the original image with a disc of radius $27h^{-1}\rm cMpc$ and $39h^{-1}\rm cMpc$, respectively.}
\label{lognormal_opening}
  \includegraphics[angle=0,width=0.8\columnwidth]{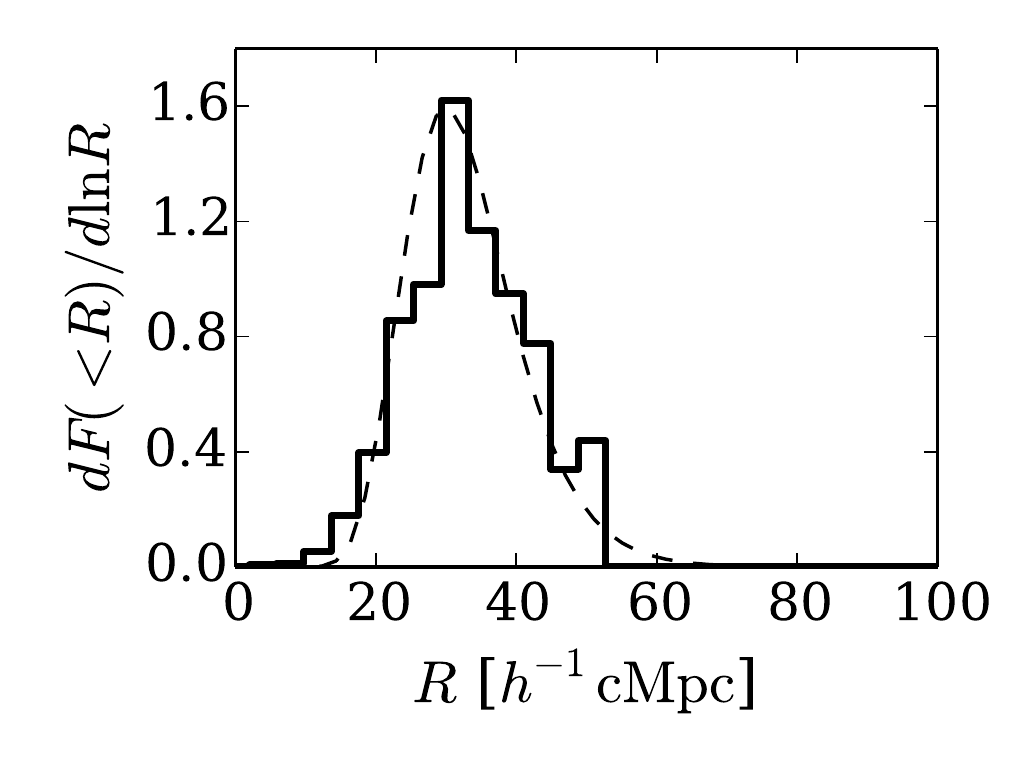}
  \vspace{-0.3cm}
  \caption{Differential size distributions of $\HII$ regions from granulometric analysis (solid line). The dashed curve shows the input probability distribution function of $\HII$ region sizes in the log-normal model.}
\label{lognormal_size_distribution}
\end{figure}

As a proof-of-concept, we apply the granulometric measurement of size distribution of $\HII$ regions to a Monte Carlo realization of the 2D log-normal bubble model from section~\ref{lognormalbubbles}. Figure~\ref{lognormal_opening} shows an example of sieving the morphology of $\HII$ regions, represented by the white areas. The original distribution of $\HII$ regions, i.e.\ the binary image $X_{\rm HII}(\mathbf{r})$, is shown in the left panel.  When it is sieved ($X\circ S_R$) with a disc
of radius $R=27h^{-1}\rm cMpc$ (middle panel), the structures smaller than the radius of the disc are removed. A larger disc of radius $R=39h^{-1}\rm cMpc$ (right panel) sieves most of the structures and only some of the largest $\HII$ regions remain.

The histogram in Figure \ref{lognormal_size_distribution} shows the differential size distribution measured with the sieving procedure. The differential size distribution clearly picks up the underlying probability distribution function  of $\HII$ region sizes used by the log-normal model (dashed curve).  
The agreement is not perfect because of overlapping $\HII$ regions, which causes a departure from spherical shapes. 
Nevertheless, the log-normal model proves that the granulometric measure of the size distribution traces the underlying probability distribution of $\HII$ regions. From this we conclude that the granulometric analysis is a very useful statistical tool to quantify the size distribution of $\HII$ regions with a mathematically well-motivated framework.

\subsection{Size distribution of $\HII$ regions: RT simulation}\label{sec:granulometry_RT}

\begin{figure}
  \centering
  \includegraphics[angle=0,width=\columnwidth]{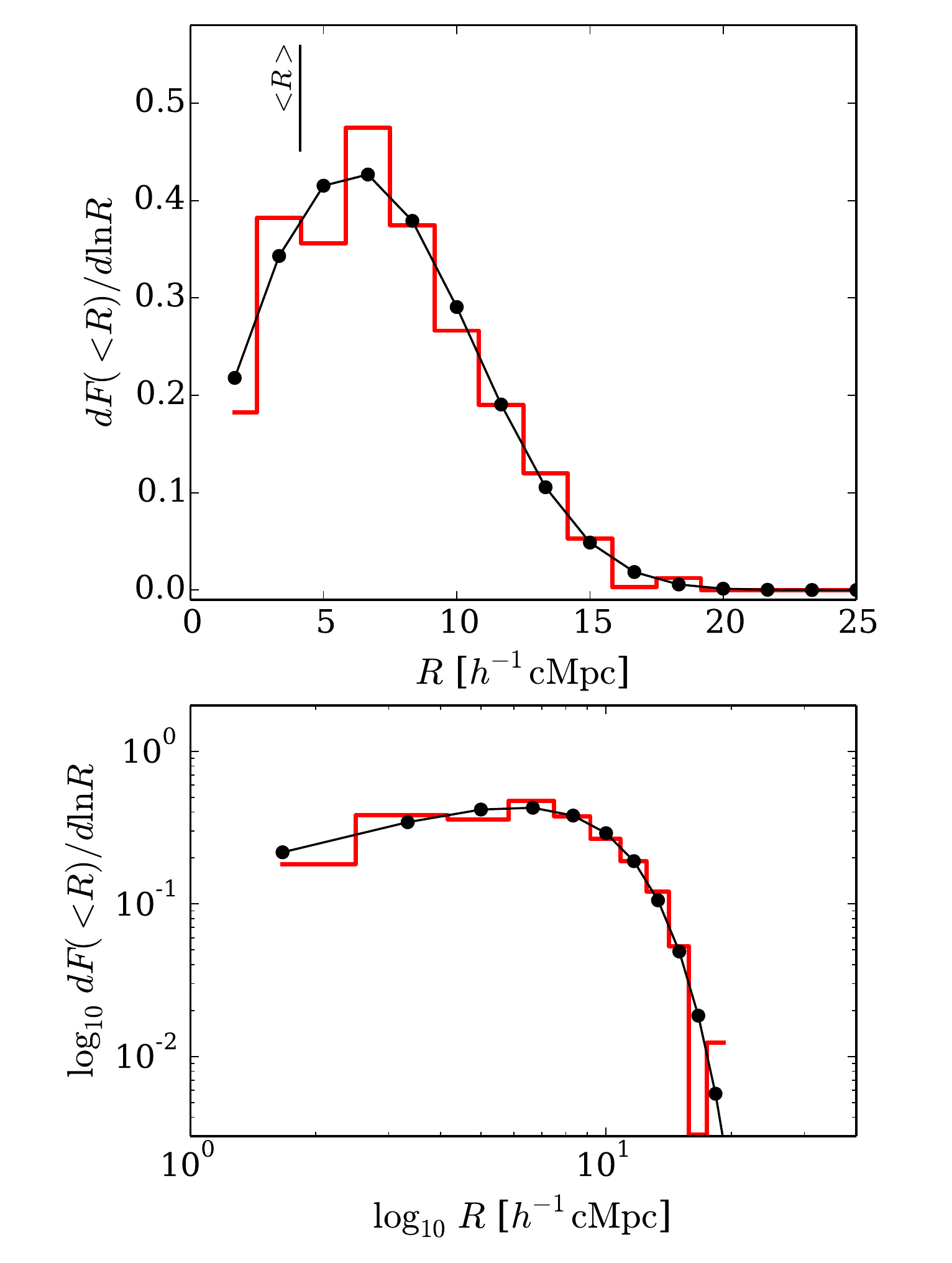}
  \caption{Differential size distribution of $\HII$ regions measured from the granulometric analysis of the ionization structure, $\xHII(\mathbf{r})$, in the RT simulation (red histograms) at $z=6.8$ ($\langle\xHII\rangle_V=0.40$). The black curves are the best-fit modified Schechter functions (see text). The vertical line indicates the mean radius of the size distribution. {\it Top}: Linear scale plot. {\it Bottom}: Log-log scale plot to show the exponential drop of the large size end and the power-law slope of the small size end of the $\HII$ regions.}
\label{pattern_spec_sim}
\end{figure}

Having demonstrated the potential of granulometry, we next apply the analysis to the RT simulation (Section \ref{sec:simulation}). Figure~\ref{pattern_spec_sim} (histogram) shows the differential size distribution of $\HII$ regions measured from the ionization fraction field $\xHII(\mathbf{r})$  at $z=6.8$. The vertical line shows the mean size of $\HII$ regions. The black curve is an analytic fitting formula (a modified Schechter function),
\begin{equation}
R\frac{dF(>R)}{dR}=\left(\frac{R}{R_0}\right)^\alpha\exp\left[-\left(\frac{R}{R_\ast}\right)^\beta\right],\label{eq:bubble_func}
\end{equation}
where $R_\ast$, $\alpha$, $\beta$ are free parameters and $R_0$ is determined from the normalization $\int_0^\infty\frac{dF(>R)}{dR}dR=1$, which gives an analytic expression $R_0=R_\ast\left[\Gamma(\alpha/\beta)/\beta\right]^{1/\alpha}$, where $\Gamma$ is the Gamma function. The granulometric measurement of the size distribution of $\HII$ regions is remarkably well described by this modified Schechter function. The position of the peak size, the power-law slope at small sizes, and the exponential cutoff at large sizes are captured perfectly. 
The best-fit parameters are $R_\ast=10.0h^{-1}\rm cMpc$, $\alpha=0.71$, $\beta=2.78$. The mean radius then has the analytic expression $\langle R\rangle=R_\ast\left[\Gamma(\frac{1+\alpha}{\beta})\right/\left.\Gamma(\frac{\alpha}{\beta})\right]\simeq 4.12h^{-1}\rm cMpc$\footnote{The mean radius of $\HII$ regions evaluated from the fitting formula differs only by 2.5\% from the direct integration of the differential size distribution measured from the simulation. As the measured median radius is also $4.18h^{-1}\rm cMpc$, we only show the mean radius of $\HII$ regions.}. We have also tested the validity of the modified Schechter function for other redshifts, when the average neutral fraction is different, and find that the size distribution is fit very well by the modified Schechter function over a wide range of redshifts. 

There are physical motivations why the $\HII$ region size distribution follows the form of a modified Schechter function. First, suppose that the UV (1500\AA) luminosity function of galaxies responsible for driving reionization follows the Schechter function $\phi(L_{1500})\propto (L_{1500}/L^\ast_{1500})^{\alpha_L}\exp\left(-L_{1500}/L^\ast_{1500}\right)$ with a characteristic luminosity $L^{\ast}_{1500}$ and a faint-end slope $\alpha_L$. The ionizing photon luminosity $\dot{N}_{\rm ion}$ (in units of $\rm s^{-1}$) at $<912$~\AA~from a galaxy is given by the product $\dot{N}_{\rm ion}=f_{\rm esc}\xi_{\rm ion}L_{1500}$ (e.g. \citealt{2013ApJ...768...71R}), where $f_{\rm esc}$ is the escape fraction and $\xi_{\rm ion}$ is the ratio between the 1500~\AA~luminosity and the intrinsic ionizing photon production rate of a galaxy. Assuming that all galaxies have their own $\HII$ regions during the early pre-overlapping phase, the bubble number density $dn_b(R)/dR$ per unit radius of $\HII$ regions is given by $\frac{dn_b(R)}{dR}dR=\phi(L_{1500})dL_{1500}$. By estimating the radius of a $\HII$ region by counting photons (cosmological Str\"omgren sphere), $R=\left[\frac{3\dot{N}_{\rm ion}t_{G}}{4\pi\bar{n}_{\rm H}(0)}\right]^{1/3}\propto L_{1500}^{1/3}$, where $t_G$ is the time interval during which a galaxy is ionizing the IGM and $\bar{n}_{\rm H}(0)$ is the comoving mean number density of hydrogen atoms. Therefore, the bubble number density scales as $dn_b(R)/dR\propto R^{3\alpha_L+2}\exp\left[-(R/R_\ast)^3\right]$, where the characteristic size $R_{\ast}$ is given by the ionizing properties of early galaxies, $R_\ast=\left[\frac{3f_{\rm esc}\xi_{\rm ion}L^\ast_{1500}t_{G}}{4\pi\bar{n}_{\rm H}(0)}\right]^{1/3}$. Hence, because $dF(<R)/dR\propto dn_b(R)/dR$, we expect the $\HII$ region size distribution to scale as $RdF(<R)/dR\propto R^{3(\alpha_L+1)}\exp\left[-(R/R_\ast)^3\right]$, as long as there is not too much overlap between bubbles. This is indeed a form of the modified Schechter function (\ref{eq:bubble_func}). A second physical motivation comes from the result of the excursion set formalism for reionization, in which the mass function for the ionized bubbles has a form of the modified Schechter function (\citealt{2004ApJ...613....1F}).

This relation between the $\HII$ region size distribution and the properties of ionizing sources allows us to provide physical interpretations for the shape of the size distribution. For smaller $\HII$ region sizes, $R<R_\ast$, the distribution scales as a power-law, $R\frac{dF(>R)}{dR}\propto R^\alpha$. Our RT simulation gives a positive slope $\alpha=0.71$, while the above simple expectation gives a negative slope $\alpha=-3$ for the faint-end slope of $\alpha_L=-2$. The positive value of the small size end of $\HII$ region size distribution can be interpreted as a result of the well-known characteristic of the reionization process, i.e. the cosmological $\HII$ regions grow by merging many small ones. Consequently, the small size end of $\HII$ regions is redistributed to larger sizes as reionization progresses. For larger $\HII$ region sizes, $R> R_\ast$, the interpretation is more complicated. The distribution shows an exponential drop off, scaling as $\propto\exp[(-R/R_\ast)^\beta]$, which could be (partially) due to the rapid decline in the population of luminous or clustered ionizing sources responsible for producing large $\HII$ regions. For a robust interpretation a future study of the relation (if there is any) between the $\HII$ region size distribution and the properties and spatial distribution of ionizing sources is required; nonetheless, one may postulate that the $\HII$ region size distribution will contain the memory of the properties of ionizing sources.

\subsection{Relation between $\HII$ regions and cold spots \newline in 21-cm tomography}\label{sec:void_bias}

Observationally we can only analyse 21-cm tomographic data. In tomographic images, the $\HII$ regions appear as cold spots, i.e.\  regions devoid of 21-cm emission. As only temperature fluctuations are observable in radio interferometry, those regions appear to have negatives values in the 21-cm images. Thus, intuitively the size distribution of cold spots could be directly translated into that of $\HII$ regions. However, a complication arises from the contribution of the density fluctuations to the 21-cm signal as low density neutral regions can also give rise to negative values in the images (see equation \ref{eq:21}). We would therefore like to answer the following question: {\it how well do the volume-filling factor and the size distribution of 21-cm cold spots trace the volume-filling factor and size distribution of $\HII$ regions?}
 
\subsubsection{Volume-filling factors of $\HII$ regions and 21-cm cold spots}\label{sec:Q}

\begin{figure}
 \centering
  \includegraphics[angle=0,width=0.9\columnwidth]{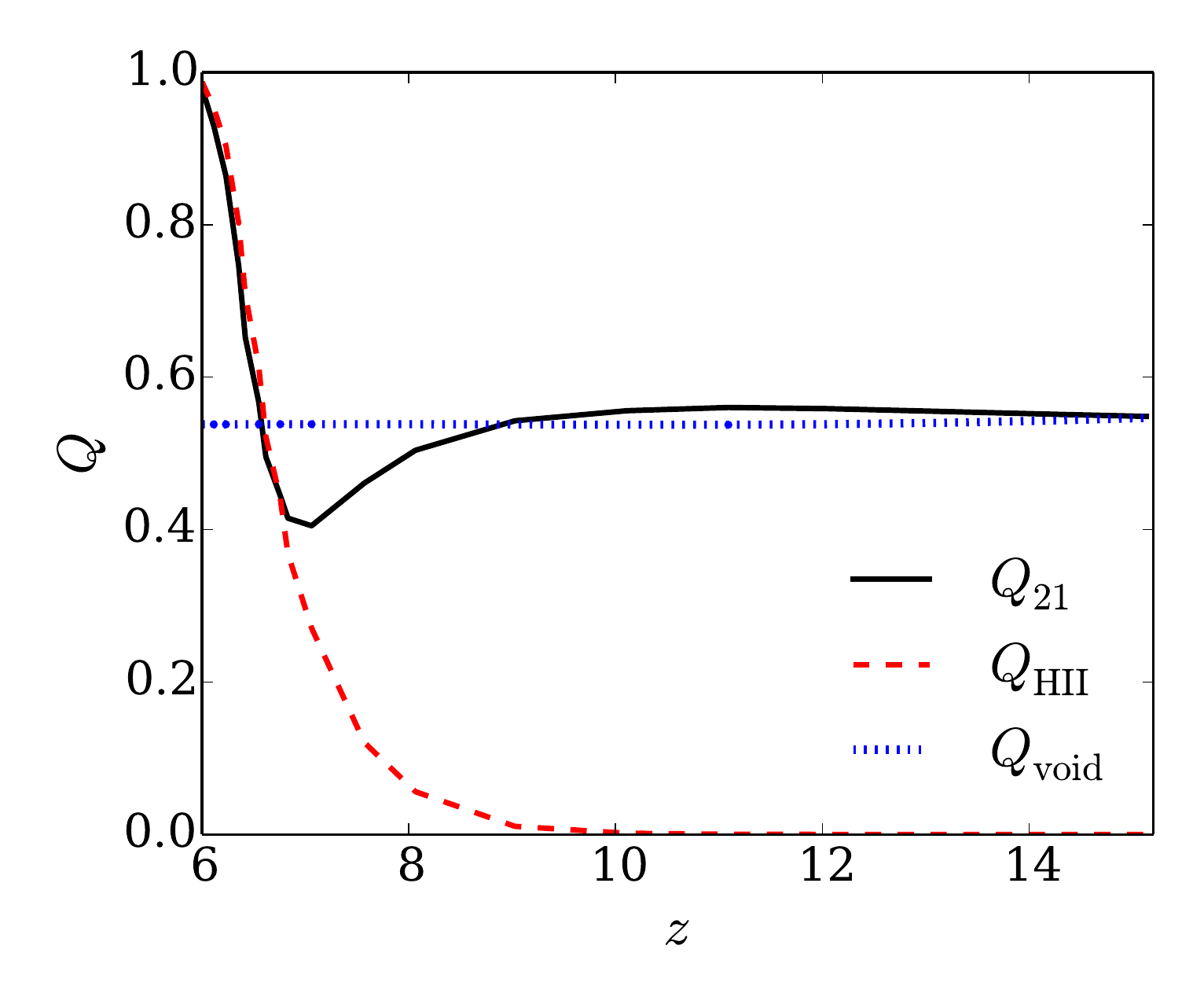}
  \caption{volume-filling factors of 21-cm cold spots (black solid line), $\HII$ regions (red dashed), and voids (blue dotted) as a function of redshift in the RT simulation.}\label{fig:Q}
\end{figure}

The 21-cm cold spots can encode the statistics of voids in matter density, as well as of ionized regions. To interpret the cold spots as a signature of $\HII$ regions we should understand the nature of the void size distribution.

Figure~\ref{fig:Q} shows the volume-filling factors of 21-cm cold spots (black solid line), $\HII$ regions (red dashed), and voids (blue dotted). The filling factors of 21-cm cold spots and $\HII$ regions are measured from the simulation box. Similarly, the filling factor of voids is calculated from the negative excursion sets $\delta_{\rm b}(\mathbf{r})<0$ (underdense regions) of the density fluctuation field.

The figure shows that in our simulation the volume-filling factor of 21-cm cold spots traces that of $\HII$ regions well for $6<z<7$. However, when the filling factor of $\HII$ regions drops below 0.4, the filling factor of 21-cm cold spots starts to deviate from that of $\HII$ regions. At $z>9$, it approaches the filling factor of voids, $Q_{\rm void}$\footnote{Note that the void filling factor $Q_{\rm void}$ is not exactly 0.5, which we naively expect from the linear perturbation theory. Because the density perturbation of the IGM is becoming mildly nonlinear during the EoR, underdense regions (voids) fill up more volume than the overdense regions. Therefore, the void filling factor is slightly larger than 0.5.}. We refer to the difficulty in interpreting 21-cm cold spots due to the contributions from both $\HII$ regions and voids as {\it the void-$\HII$ regions confusion}. 

The transition around a filling factor of 0.4 can be understood as follows. The level of the void contamination is controlled by the probability that the negative excursion of density fluctuation passes below the threshold given by the mean 21-cm signal $\langle\delta T_{21}\rangle=T_0\langle \xHI(1+\delta_{\rm b})\rangle\approx T_0(1-Q_{\rm HII})$. Because the 21-cm signal from the neutral parts of the IGM is $\delta T_{21}=T_0(1+\delta_{\rm b})$, in order for the density fluctuations to appear as cold spots in the mean subtracted (observed) 21-cm signal, i.e.\ $\Delta T_{21}=\delta T_{21}-\langle\delta T_{21}\rangle<0$, the density fluctuations must be lower than the threshold value, $\delta_{\rm b}<-Q_{\rm HII}$. The rms density fluctuations in our simulation are $\sigma_{\rm b}\approx0.37[(1+z)/8]^{-1}$. The volume-filling factor of $\HII$ regions $Q_{\rm HII}\approx0.4$ corresponds to this rms density fluctuation level. Below this value of $Q_{\rm HII}$, we expect a large contamination from the density fluctuations to 21-cm cold spots since many $1\sigma_{\rm b}$ fluctuations can pass below the threshold level $\approx -Q_{\rm HII}$. This is the reason why the assumption that 21-cm cold-spots filling factor is a good tracer of $\HII$ regions breaks down around $Q_{\rm HII}\approx0.4 (\approx\sigma_{\rm b})$. 

However, this is not a fundamental limitation of the granulometric analysis of tomographic data; the apparent confusion is associated with the somewhat arbitrary choice of the threshold value $\Delta T^{\rm th}_{21}=0$ when creating a binary field. In a future improved granulometric analysis the use of multiple threshold values should allow us to mitigate the apparent confusion to a large extent. In this introductory paper on granulometry however, we adopt a single threshold value $\Delta T^{\rm th}_{21}=0$. In the following, we discuss the factors controlling the void-$\HII$ region confusion to help its mitigation in future work.

\subsubsection{Size distributions of $\HII$ regions and 21-cm cold spots}

\begin{figure}
 \begin{center}
  \includegraphics[angle=0,width=\columnwidth]{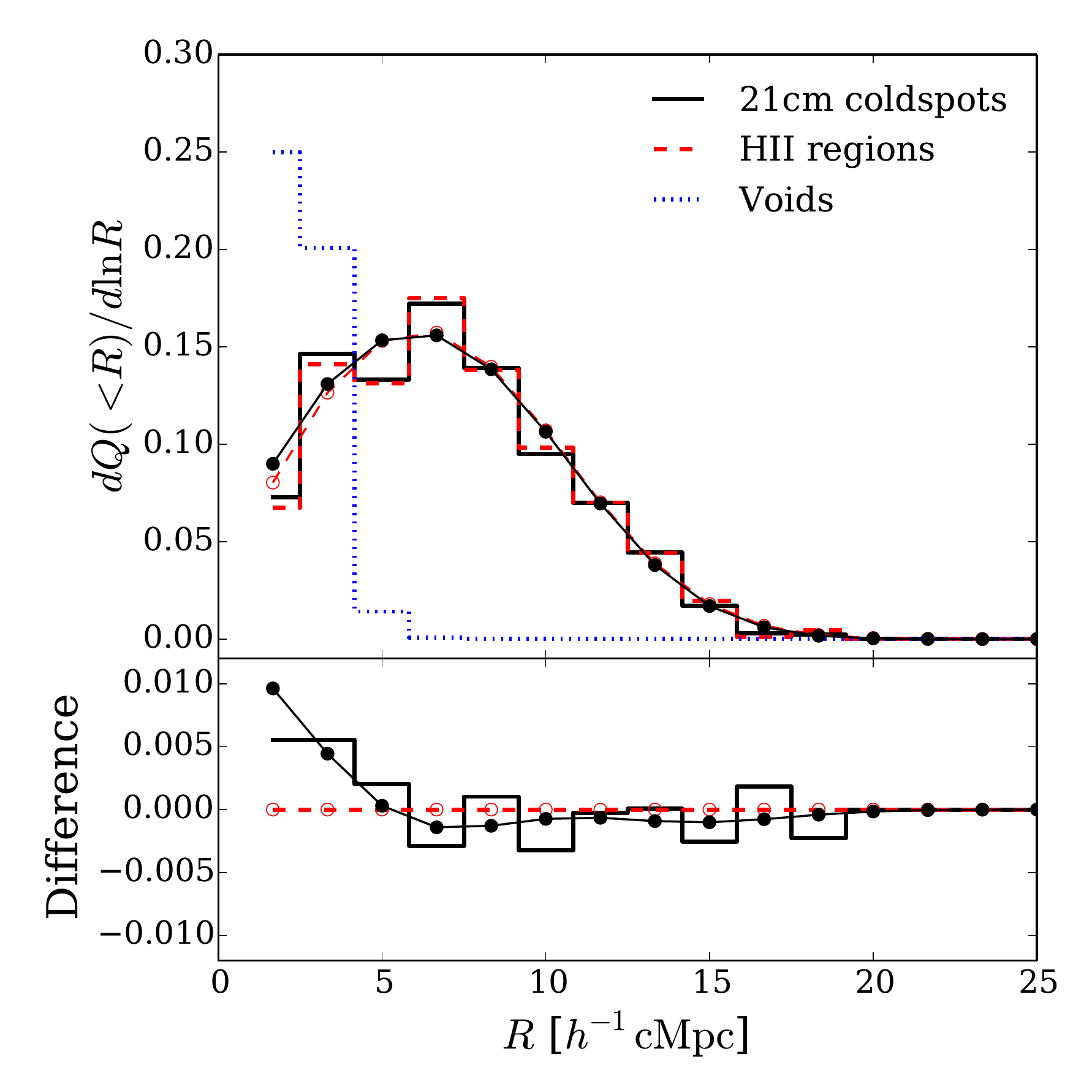}
  \caption{{\it Top}: Size distribution of 21-cm cold spots (black solid histogram), $\HII$ regions (red dashed histogram), and voids (blue dotted histogram) from the granulometric analysis of the RT simulation at $z=6.8$ ($\langle\xHII\rangle_V=0.40$). The black and red curves with filled and open circles are the best-fit modified Schechter functions for the size distributions of 21-cm cold spots and $\HII$ regions, respectively. {\it Bottom}: Difference between the size distributions of 21-cm cold spots and $\HII$ regions, $dQ_{21}(<R)/d\ln R-dQ_{\rm HII}(<R)/d\ln R$  (black). The red line is the line of zero difference.}\label{fig:cold spot}
 \end{center}
\end{figure}

Figure \ref{fig:cold spot} (top panel) shows the size distributions of 21-cm cold spots $dQ_{21}(<R)/dR$, $\HII$ regions $dQ_{\rm HII}(<R)/dR$, and voids $dQ_{\rm void}(<R)/dR$ measured by the granulometric analysis of the RT simulation at $z=6.8$. At this redshift, the void-$\HII$ region confusion is small. The size distribution of voids is measured by applying the same granulometric analysis to the negative excursion sets of the density fluctuation field, $\delta_{\rm b}(\mathbf{r})<0$.  The bottom panel shows the difference between the size distributions of 21-cm cold spots and $\HII$ regions. Note that the size distributions shown in Figure \ref{fig:cold spot} (and hereafter unless otherwise stated) are normalized to the volume-filling factor, i.e. $dQ(<R)/dR\equiv QdF(<R)/dR$\footnote{This definition is more convenient when there is contamination from voids. Normalizing the size distribution to unity produces an artificial difference between the size distributions of 21-cm cold spots and $\HII$ regions at larger sizes $R$. While the void contamination is mostly confined at smaller $R$, when normalized to unity the size distribution at larger $R$ appears to be lower to compensate for the increase at small $R$. This trivial error is avoided by normalizing the size distribution to the volume-filling factor.}.

As suggested by the results on the filling factors in the previous section, the size distribution of 21-cm cold spots traces that of $\HII$ regions very well for $Q_{\rm HII}>0.4$. The slight deviation seen at $R\lesssim 5h^{-1}\rm cMpc$ occurs because of the contamination by voids. This leads to a slight underestimation of the mean size of $\HII$ regions inferred from the mean size of 21-cm cold spots (by $\sim10\%$). The voids remain a minor contaminant to the overall 21-cm cold-spot size statistics of 21-cm tomography affecting the results only by approximately 10\%.

\begin{figure}
 \begin{center}
  \includegraphics[angle=0,width=\columnwidth]{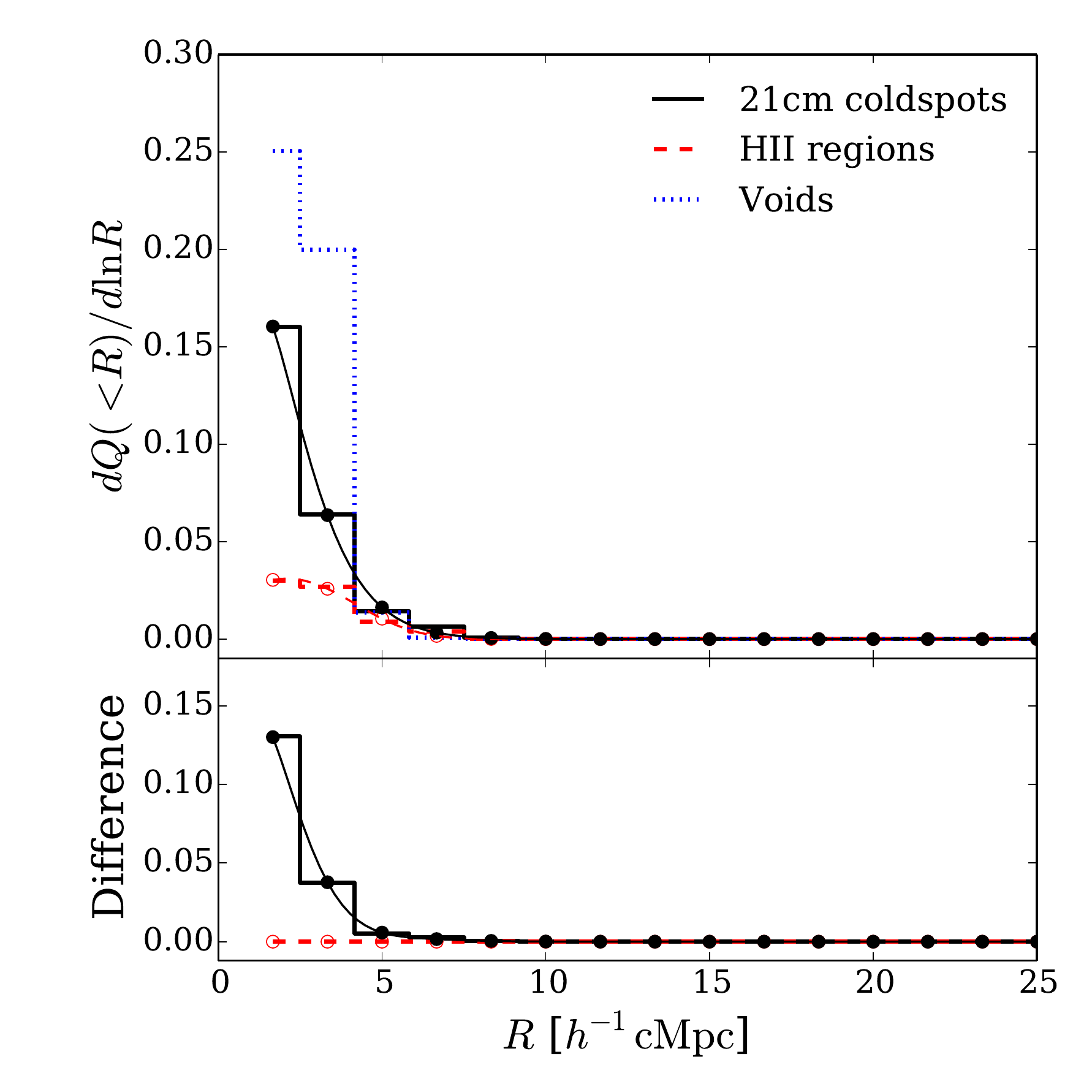}
  \caption{Same as Figure~\ref{fig:cold spot}, but at $z=7.6$ ($\langle\xHII\rangle_V=0.16$). Note that the void size distribution is virtually identical to the one at $z=6.8$ shown in Figure~\ref{fig:cold spot} because the density perturbation evolves very slowly. The void contamination is large at this redshift.}\label{fig:cold spot2}
 \end{center}
\end{figure}

Figure~\ref{fig:cold spot2} is the equivalent of Figure~\ref{fig:cold spot} at $z=7.6$ ($\langle\xHII\rangle_V=0.16$). As expected, we see that the contamination from voids dominates over the signature of $\HII$ regions in the 21-cm cold-spot size distribution. As noted in Section~\ref{sec:Q} this is because the underlying filling factor of $\HII$ regions is small during the first half of reionization.

The large difference between the shapes of the size distributions of $\HII$ regions and voids provides a way to avoid the void-$\HII$ region confusion. The void size distribution is always confined within $R\lesssim5h^{-1}\rm cMpc$. In fact, the size distribution of the excursion sets of a Gaussian random field can be well understood and predicted given a priori knowledge of the matter power spectrum (\citealt{1986ApJ...304...15B, 1987MNRAS.226..655B,2004MNRAS.350..517S}). Therefore, we can test the robustness of the identification of cold spots as $\HII$ regions against the null hypothesis of void size statistics. 

Overall, we conclude that the size distribution of 21-cm cold spot traces that of $\HII$ regions very well during the second half of reionization. For the first half of reionization the interpretation of 21-cm cold spots becomes increasingly difficult because of large contamination from voids for our canonical choice of threshold $\Delta T_{12}^{\rm th}=0$. We note that this may be mitigated by choosing lower values for the threshold.

\section{Recovery of $\HII$ region size distribution: 3D data sets}\label{sec:result_3d_granulometry}

So far we have only considered the case of a pure simulated 21-cm signal. We now turn our attention to the prospects for recovering the $\HII$ region size distribution from 21-cm tomography using a SKA-like radio interferometer.

The recovery of the size distributions of 21-cm cold spots and $\HII$ regions from real-world radio interferometric observations is subjected to noise, instrumental response, foregrounds, and observing strategy. We therefore ask the question: {\it what are the requirements to recover the $\HII$ region size distribution from 21-cm tomographic data observed with the SKA?}

\begin{figure}
  \centering
  \includegraphics[angle=0,width=\columnwidth]{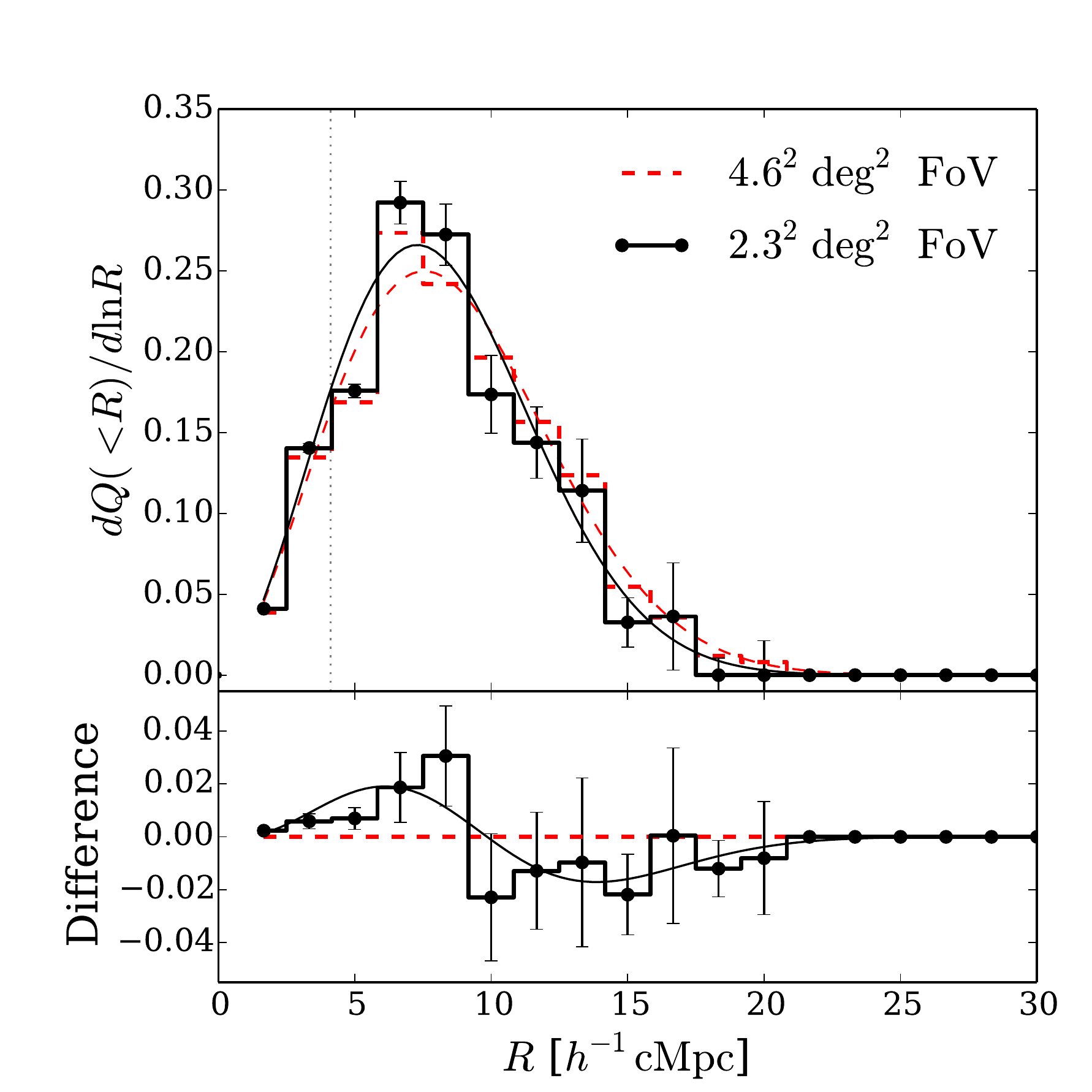}
  \caption{Effect of a finite FoV. {\it Top}: 21-cm cold-spot size distributions measured from the mock SKA1-low noise-free image cubes of $4.6^2\rm~deg^2$ FoV (red dashed histogram) and $2.3^2\rm~deg^2$ FoV (black solid histogram) at $z=6.8$ ($\langle\xHII\rangle_V=0.40$). The $1\sigma$ error due to the sample variance is shown (Section~\ref{sec:error}). The black solid and red dashed curves are the best-fit modified Schechter functions. The vertical dotted line indicates the angular resolution ${\rm FWHM}=4h^{-1}\rm cMpc$ (max. baseline 2 km). {\it Bottom}: Absolute difference, $dQ(>R)/d{\rm ln}R|_{\rm 2.3^2deg^2}-dQ(>R)/d{\rm ln}R|_{\rm \rm 4.6^2deg^2}$, between the two FoVs.}
\label{fig:FoV_effect}
\end{figure}

\subsection{Effect of the field-of-view: sample variance error}

The first requirement addresses the field-of-view (FoV) or sky coverage.
Figure~\ref{fig:FoV_effect} shows the effect of a finite FoV on the observed size distributions of 21-cm cold spots measured from the granulometric analysis of our mock image cube of $4.6^2\rm~deg^2$ FoV with frequency width 44~MHz (red dashed lines), and $2.3^2\rm~deg^2$ FoV with frequency width 22~MHz (black solid), using the SKA1-low with the limiting case of noise-free data. The frequency width is chosen to match with the size of the FoV so that the comoving lengths of the line-of-sight and perpendicular directions are equal. The error bars correspond to $1\sigma$ uncertainty due to the sample variance (Section~\ref{sec:error}). The red and black curves are the best-fit modified Schechter functions, and the vertical dotted line indicates the FWHM of the angular resolution. Note however that the cubes we have analysed have a large frequency width which implies that the real data will be subject to the light cone effect \citep{2006MNRAS.372L..43B,2012MNRAS.424.1877D}. The results in this section should therefore be interpreted as four (one) pointed observations of a FoV of  $4.6^2\rm~deg^2$ ($2.3^2\rm~deg^2$) with a frequency width of approximately 10~MHz, for which the light cone effect can most likely be neglected \citep{2014MNRAS.442.1491D}.

The figure shows that the smaller FoV ($2.3^2\rm~deg^2$) does not introduce any significant systematic bias compared to our fiducial FoV of $4.6^2\rm~deg^2$. The slight under(over)estimation of the large(small) sizes of 21-cm cold spots remains within $10\%-20\%$ at most, which is within the $1\sigma$ error due to sample variance. The sample variance error increases for larger 21-cm cold spot sizes because the number of large $\HII$ regions within a FoV is smaller. Note that below the angular resolution scale the error appears to be negligibly small. This is because the angular smoothing artificially induces any cold spots at that scale to have always the same size, and therefore the FoV-to-FoV variation is suppressed.

No systematic bias will be introduced as long as the FoV is large enough to include the largest scale of $\HII$ regions (the theoretical prediction for the characteristic size is of order of tens of comoving Mpc, e.g. $\sim0.5\rm~deg$ for $\sim50h^{-1}\rm~cMpc$. See for example \citealt{2004Natur.432..194W,2015arXiv151200564G,2016MNRAS.456.3011D}).  Therefore, the single pointing FoV of SKA1-low ($\sim3\rm~deg$) is large enough to measure the 21-cm cold-spot size distribution with a reasonable sample variance error when the 3D image cube is used.

\begin{figure*}
  \centering
  \includegraphics[angle=0,width=\columnwidth]{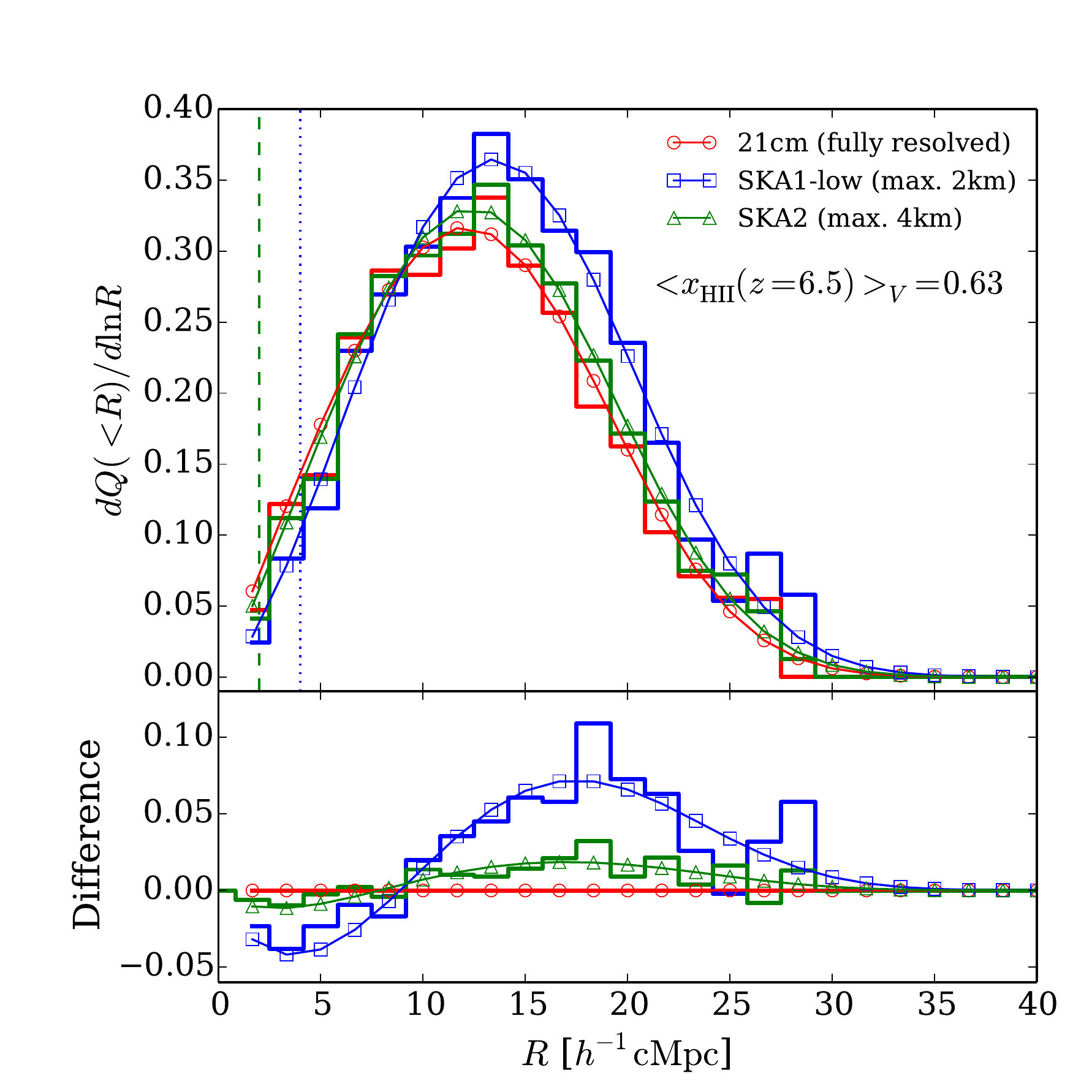}
  \includegraphics[angle=0,width=\columnwidth]{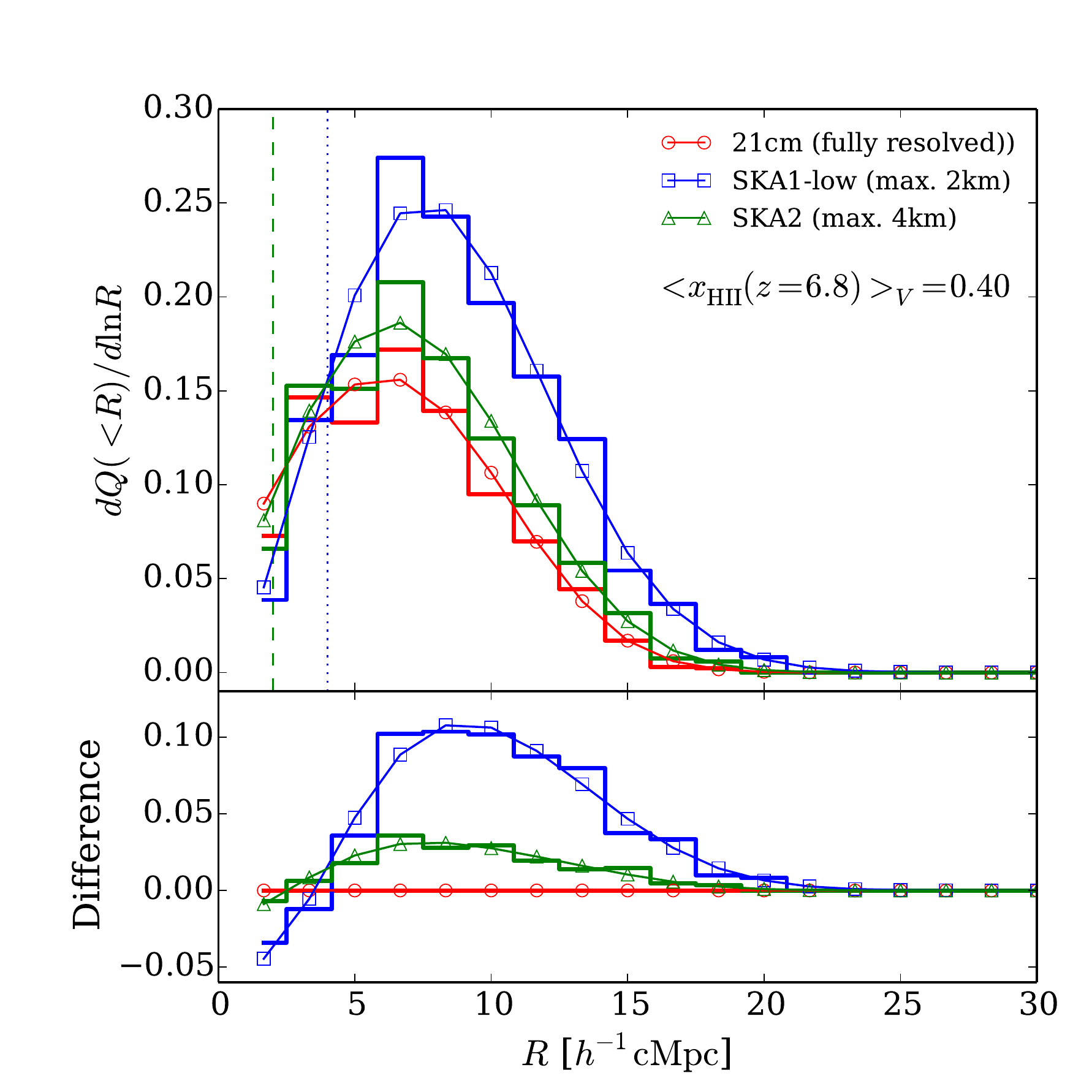}
  \caption{{\it Top}: 21-cm cold-spot size distribution with the angular resolution of SKA1-low (max. baseline 2 km, blue histogram) and SKA2 (max. baseline 4 km, green histogram) at $z=6.5$ (left panel) and 6.8 (right panel). The red histogram shows the case of a fully resolved 21-cm signal. The red, blue, green curves with open circles, triangles, and squares show the best-fit modified Schechter functions. The vertical lines show the angular resolutions of SKA1-low (2.8 arcmin, blue dotted) and SKA2 (1.4 arcmin, green dashed) in comoving length. {\it Bottom}: Absolute difference, ${\rm abs.~error}=dQ(>R)/d{\ln}R|_{\rm data}-dQ(>R)/d{\ln}R|_{\rm sim}$, between the data models and the simulation. The figure illustrates that the smoothing that results from a finite angular resolution merges small into large cold spots.}
\label{resolution_effect}
\end{figure*}

\subsection{Effect of angular resolution: smoothing bias}\label{sec:mixing_bias}

The second requirement comes from the finite angular resolution of the instrument, which strongly affects the recovery and interpretation of the 21-cm cold-spot and $\HII$ region size distributions. 

Figure \ref{resolution_effect} shows the effect of finite angular resolution on the 21-cm cold-spot size distribution at redshift $z=6.5$ and $6.8$ for a noiseless image cube with a $4.6^2\rm~deg^2$ FoV and a frequency depth of 44~MHz, using an SKA1-low (blue squares) and SKA2 (green triangles) configuration. The frequency resolution is set to match the angular resolution in comoving coordinates. The size distribution of the pure simulated 21-cm tomography (red circles) represents the case of an ideal, fully resolved observation. The vertical lines indicate the angular resolutions of SKA1-low (2.8 arcmin, blue dotted) and SKA2 (1.4 arcmin, green dashed).

The angular resolution introduces a systematic bias in the observed size distribution of 21-cm cold spots. Both at $z=6.5$ and $6.8$, the shape of the size distributions is systematically shifted towards larger sizes. This is somewhat counter-intuitive as an obvious expectation is the suppression of the small end of a size distribution. This systematic bias occurs because the angular smoothing (or Gaussian PSF) actually mixes and merges many 21-cm cold spots (including both $\HII$ regions and voids). As a result, many small cold spots become a large one. We call this important effect the {\it smoothing bias}. Since the granulometric analysis is performed on the binary field satisfying $\Delta T_{21}<0$, smoothing does not always remove small cold spots. Instead, because of the mixing of cold spots, intrinsically small cold spots re-populate the larger part of a size distribution. Quantitatively, the impact of the smoothing bias is shown in the bottom panels of Figure~\ref{resolution_effect}, where the difference between the simulation and the mock observations is plotted. The smoothing bias is larger for a larger angular smoothing scale. Importantly, the figure shows that {\it the smoothing bias impacts all scales of the size distribution, even when the angular resolution scale is below the scale of interest}.

The impact of a finite angular resolution is less severe when the characteristic size or mean size of the underlying (true) 21-cm cold-spot size distribution is much larger than the angular resolution, $\theta_{\rm A}\ll R_\ast/D_A(z)$ [or $\ll\langle R\rangle/D_A(z)$], where $D_A(z)$ is the angular diameter distance to redshift $z$. For example, at $\langle x_{\rm HII}\rangle_V=0.40$ the angular resolution causes a systematic bias in size distribution by $\sim 50(20)\%$ fractional error for SKA1-low (SKA2), while at $\langle x_{\rm HII}\rangle_V=0.63$ the bias is only about $\sim 10\%$. We note that a chromatic PSF will introduce an extra complication (\citealt{2012ApJ...745..176V}) and possibly a systematic bias. However, addressing this is beyond the scope of this paper.

This analysis suggests that the effect of angular resolution must be taken into account when interpreting the 21-cm cold-spot size distribution. If an unbiased measurement (to the accuracy of order 10\%) of the size distribution is required, we should correct the angular resolution bias by a: (i) `baseline-design' approach, where we ensure that the maximum baseline is large enough that the angular resolution bias in the size distribution and the characteristic size of cold spots remains sufficiently small, or a (ii) `forward-modelling' approach, where we create a large suite of models and mock data cubes and directly fit the simulated size distribution in the observation using a Markov Chain Monte Carlo approach. However, as we will show in Section~\ref{sec:discussions}, a sufficiently high angular resolution must be achieved for both methods to work.

\begin{figure*}
 \begin{center}
 \includegraphics[angle=0,width=0.9\textwidth]{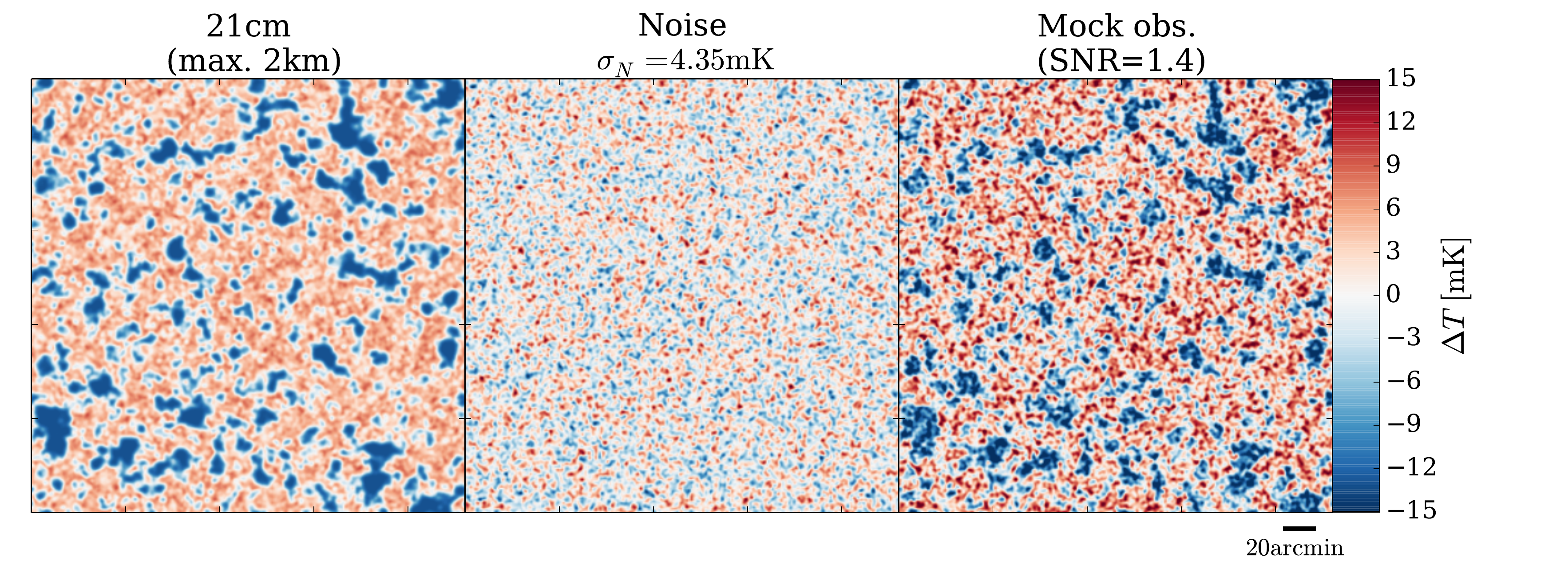}
  \caption{Mock 2D images at $z=6.8$ as observed by SKA1-low with $\sigma_{\rm N}=4.35\rm~mK$ rms noise level. The left, middle, and right panels show the image of the noiseless 21-cm signal, noise, and mock data with a $\rm SNR=1.4$. All the images have a FoV of $4.6\times4.6\rm~deg^2$ ($500h^{-1}\rm cMpc$ on a side).}
\label{fig:noise_map}
 \end{center}
\end{figure*}

\begin{figure*}
 \begin{center}
  \includegraphics[angle=0,width=\columnwidth]{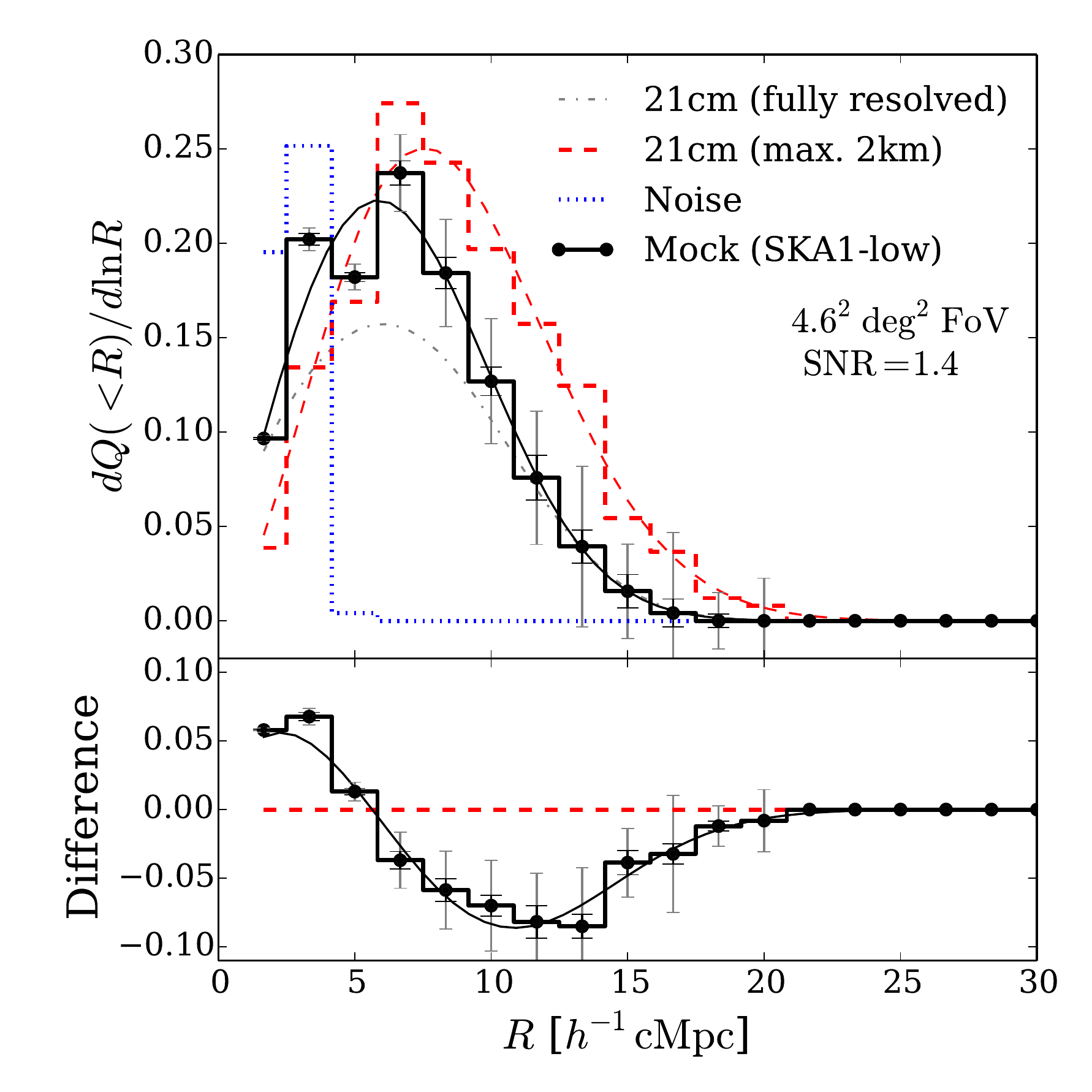}
  \includegraphics[angle=0,width=\columnwidth]{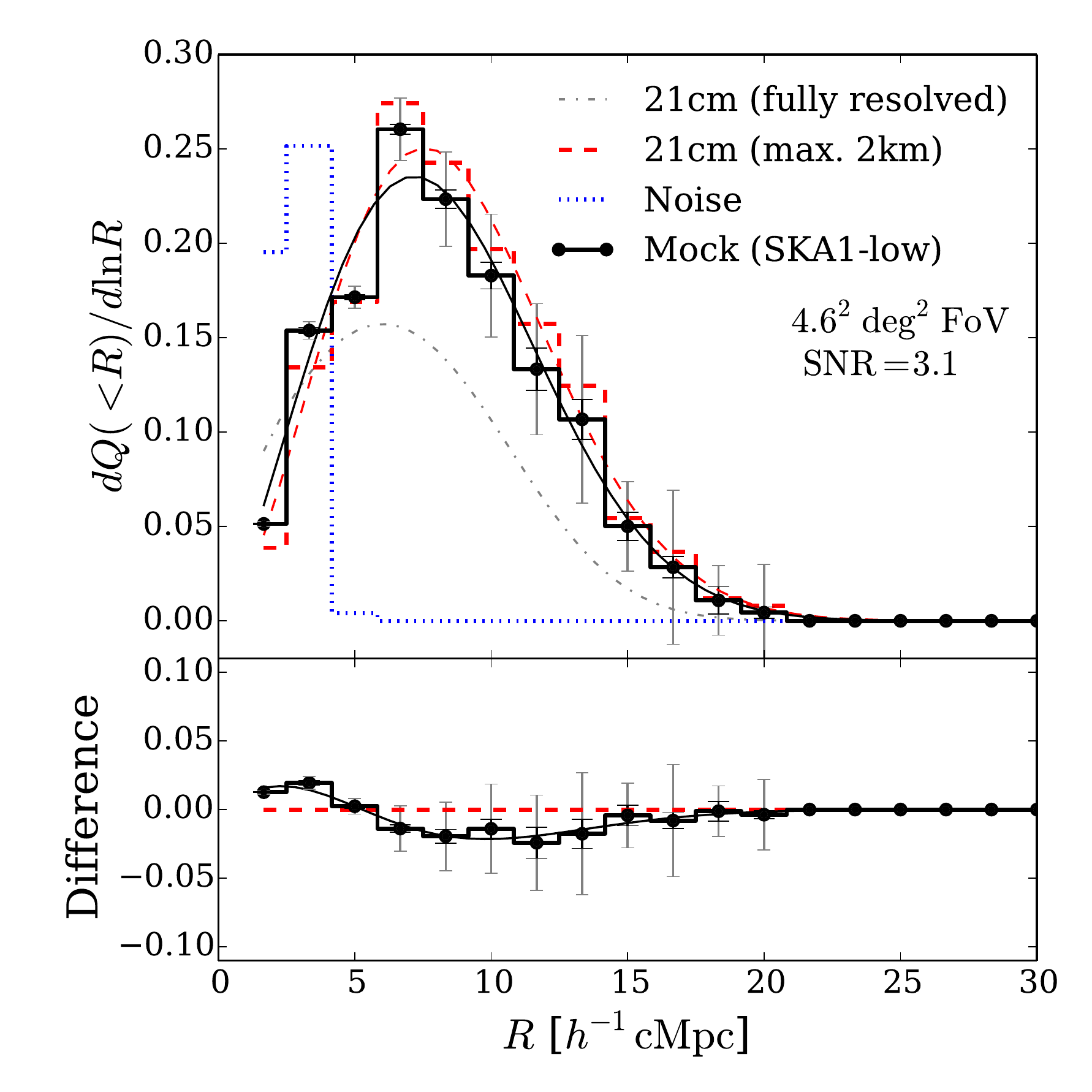}
  \caption{{\it Top}: 21-cm cold-spot size distributions measured from the pure 3D 21-cm data cube with a 2 km baseline angular resolution (red dashed), noise cube (blue dotted), and mock SKA1-low observation with $\sigma_{\rm N}=4.35\rm~mK$ (left panel) and $\sigma_{\rm N}=2.0\rm~mK$ (right panel) rms noise level at $2.3\rm~arcmin$ resolution. The curves show the best-fit modified Schechter functions. {\it Bottom}: Absolute difference between the pure 21-cm signal and mock observations. The inner error bars indicate the $3\sigma$ statistical uncertainties due to the thermal noise. The outer error bars include $3\sigma$ statistical error from the sample variance (see Section~\ref{sec:error}). The grey dotted-dash curve is the best-fit modified Schechter function in the fully resolved 21-cm signal in the RT simulation.}\label{fig:noise_effect}
 \end{center}
\end{figure*}

\subsection{Effect of thermal noise: splitting bias}\label{sec:noise_effect}

The third requirement is on the thermal noise. Figure~\ref{fig:noise_map}, showing a 2D image from the data cube, visually illustrates the effect of noise on the cold spots in 21-cm tomography. The thermal noise (middle panel) acts as an additive noise, and contaminates the underlying 21-cm cold spots (left panel) by distorting the shape and size of the $\HII$ regions (right panel).                

The effect of thermal noise on the observed 21-cm cold-spot size distribution is shown in Figure~\ref{fig:noise_effect}, which plots the measured size distributions from a mock 3D image cube with SKA1-low (21-cm+noise, black solid), noiseless 21-cm signal (21-cm, red dashed), and noise cube (noise, blue dotted). The $3\sigma$ error due to the thermal noise (inner error bars) and the total uncertainty including the $3\sigma$ sample variance error (outer error bars) are shown. The left (right) panel corresponds to the two mock observations with SKA1-low having an rms noise level of $4.35\rm~mK$ ($2.0\rm~mK$) per resolution element for an interferometer with a 2 km maximum baseline. 

First, as shown in the left panel in Figure~\ref{fig:noise_effect}, a SKA1-low with $\sigma_{\rm N}=4.35\rm~mK$ permits us to measure the 21-cm cold-spot size distribution using a low SNR 3D image cube (black solid) within $\lesssim50\%$ of the noiseless case (red dashed). 

The thermal noise, however, introduces another important systematic bias. The size distribution measured from the low SNR image cube is shifted systematically towards lower sizes beyond the statistical error. This bias occurs because the noise splits up the cold spots into smaller ones when the positive excursion sets of the noise fluctuation occur inside 21-cm cold spots. We thus call this systematic bias the {\it splitting bias}. The bottom panels in Figure~\ref{fig:noise_effect} indeed show that originally large 21-cm cold spots are split and re-populate the small end of the size distribution. 

\begin{figure}
 \hspace{-3mm}
  \includegraphics[angle=0,width=1.05\columnwidth]{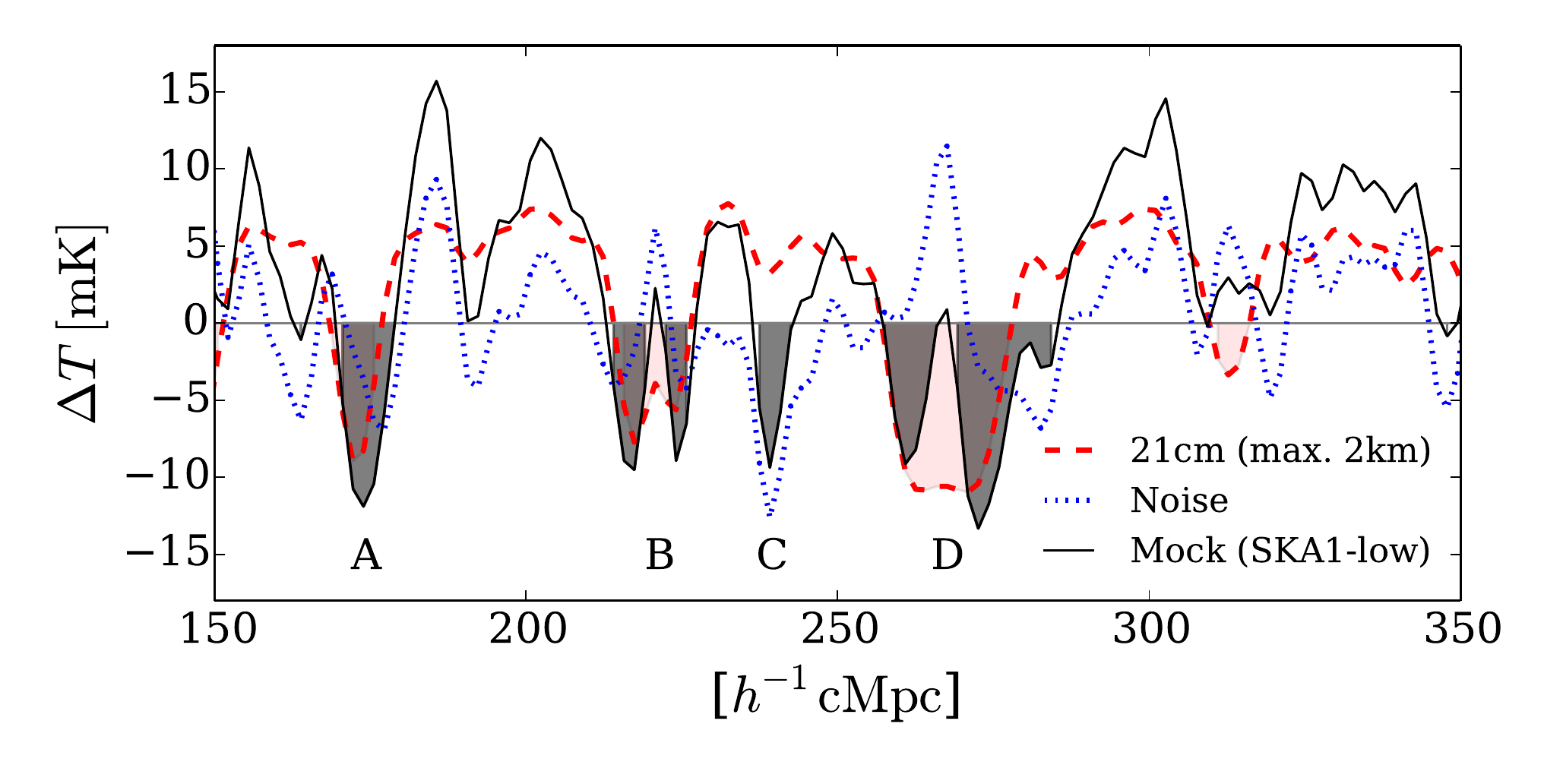}
  \vspace{-0.7cm}
  \caption{Origin of a splitting bias due to thermal noise. The 1D profiles of the brightness temperature contrast of pure 21-cm signal (red dashed line), noise (blue dotted), and mock data, i.e. 21-cm+noise (black solid) along the perpendicular direction on the sky. The true (red shaded) and observed (black shaded) cold spots are marked as shaded regions. We label the four example 21-cm cold spots with A, B, C, and D for the ease of discussion.}\label{fig:splitting_bias}
\end{figure}

To understand the mechanism of the splitting bias, Figure~\ref{fig:splitting_bias} shows the 1D profiles of the brightness temperature contrast for the same three data sets as above. The cold spots B and D are split into two separate ones in the mock observation. This occurs when the positive excursion sets of noise exceed the contrast of the underlying 21-cm cold spots, $\Delta T_{\rm coldspot}\approx5-10\rm~mK$. Because the noise is a Gaussian random field, the probability that the positive excursion of noise exceeds the 21-cm cold spot contrast is $P(>\Delta T_{\rm coldspot})=\frac{1}{2}\left[1-{\rm erf}\left(\frac{\Delta T_{\rm coldspot}}{\sqrt{2}\sigma_{\rm N}}\right)\right]\approx1-12\%$ for $\Delta T_{\rm coldspot}\approx5-10\rm~mK$, where ${\rm erf}$ is the error function. This is a noticeable effect for $\sigma_{\rm N}=4.35\rm~mK$ (as shown in the left panel of Figure~\ref{fig:noise_effect}) where the $\sim2\sigma$ peaks of noise contaminate the signal inducing the splitting bias. On the other hand, this means that only high sigma peaks act as contaminants to the granulometric measurement of the size distribution. This is the reason why a low SNR image cube still permits a reasonable recovery of the underlying size distribution of 21-cm cold spots.

Note that an opposite effect also occurs: a creation of spurious cold spots due to the negative excursion sets of noise (for example see the cold spot C in Figure~\ref{fig:splitting_bias}). However, because these spurious cold spots are always small ($R\lesssim 5h^{-1}\rm~cMpc$), the net effect of noise is a systematic bias shifting the size distribution towards smaller sizes.

At a noise level of $\sigma_{\rm N}=2.0\rm~mK$ (i.e.\ $\rm SNR=3.1$), shown in the right panel of Figure~\ref{fig:noise_effect}, the splitting bias becomes much smaller. The probability that contaminating noise peaks occurs for 21-cm cold spots with $\Delta T_{\rm coldspot}\approx5-10\rm~mK$ is only $P(>\Delta T_{\rm coldspot})<1\%$.

The level of the systematic bias entering into the cold-spot size distribution measurement does not scale linearly with the noise level (and hence, with the integration time). There is a jump in the quality of the recovery of the 21-cm cold-spot size distribution as a function of the rms noise/integration time. When the rms noise passes the threshold required by the expected brightness temperature contrast of 21-cm cold spots, the quality of the measurement jumps up. Above this threshold, increasing the integration time gradually improves the quality, but the gain is not as pronounced.

The statistical uncertainty of the thermal noise is negligibly small and is shown as the small inner error bars in Figure~\ref{fig:noise_effect}. This is because in a large FoV ($4.6\times4.6\rm~deg^2$) the variation of cold spot sizes due to the thermal noise tends to average out. As the large end of the size distribution has fewer cold spots to average out the noise error, the statistical error bars increase slightly at larger sizes. On the other hand, the error bars become negligibly small for smaller cold spots as there are many samples.

\begin{figure}
 \begin{center}
  \includegraphics[angle=0,width=\columnwidth]{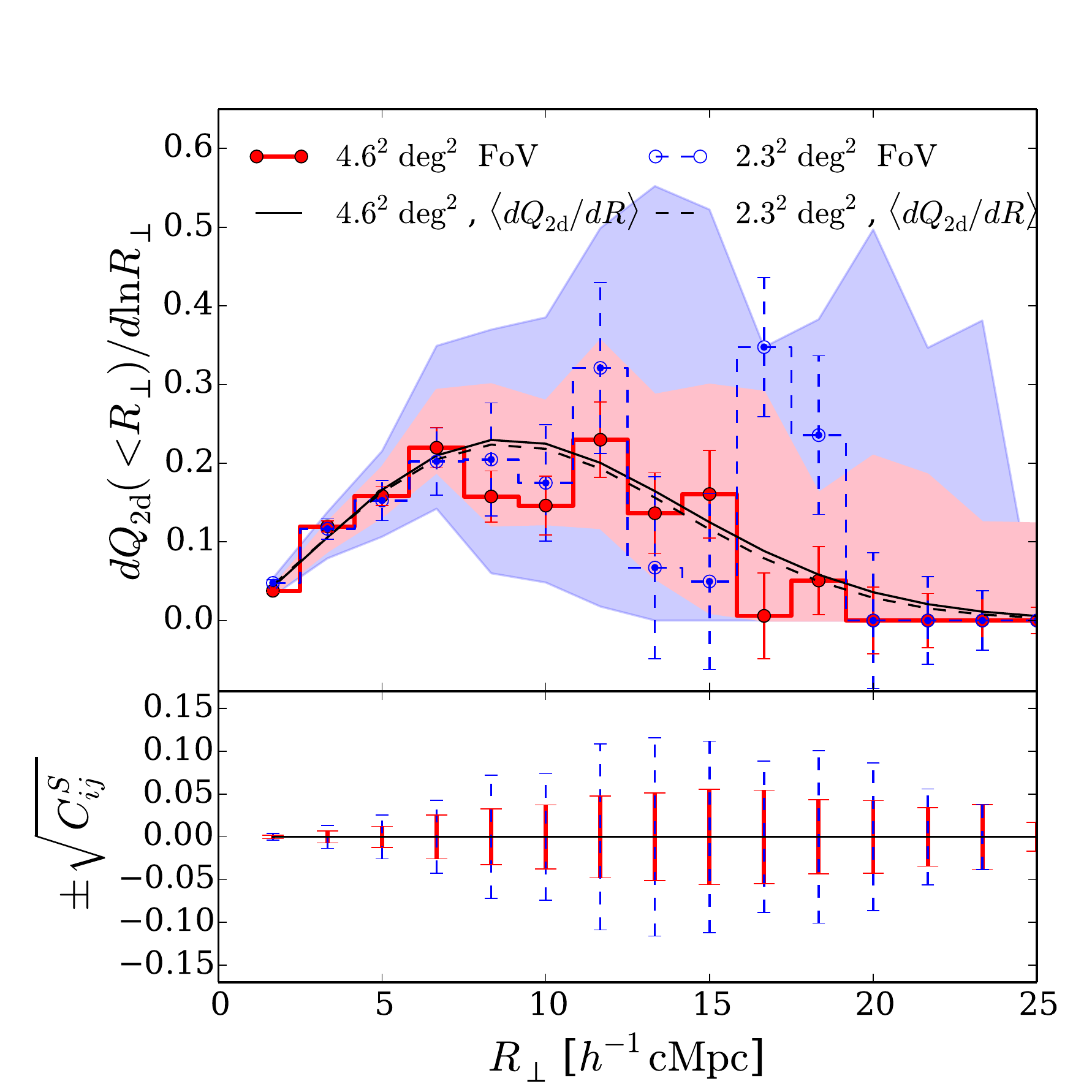}
  \caption{Effect of a finite FoV on the measured 2D size distribution from a 21-cm image in mock tomographic data. {\it Top panel}: The red solid (blue dashed) histogram shows the 21-cm cold-spot size distribution measured from a 2D image slice with $4.6^2\rm~deg^2$ FoV ($2.3^2\rm~deg^2$ FoV) from SKA1-low at noiseless limit. The $1\sigma$ error bars due to the sample variance are shown. The red (blue) shaded regions brackets the maximum and minimum values appeared in 100 random images with $4.6^2\rm~deg^2$ FoV ($2.3^2\rm~deg^2$ FoV). The black solid (dashed) curve shows the best-fit modified Schechter function to the mean of the size distributions measured from 100 random images. {\it Bottom panel}: Magnitude of the sample variance uncertainty determined from the covariance matrix. The figure shows that a larger FoV (or multiple FoVs) is preferred to measure the 21-cm cold-spot size distribution reliably.}
\label{fig:FoV_effect_2D}
 \end{center}
\end{figure}

\begin{figure*}
 \begin{center}
  \includegraphics[angle=0,width=\columnwidth]{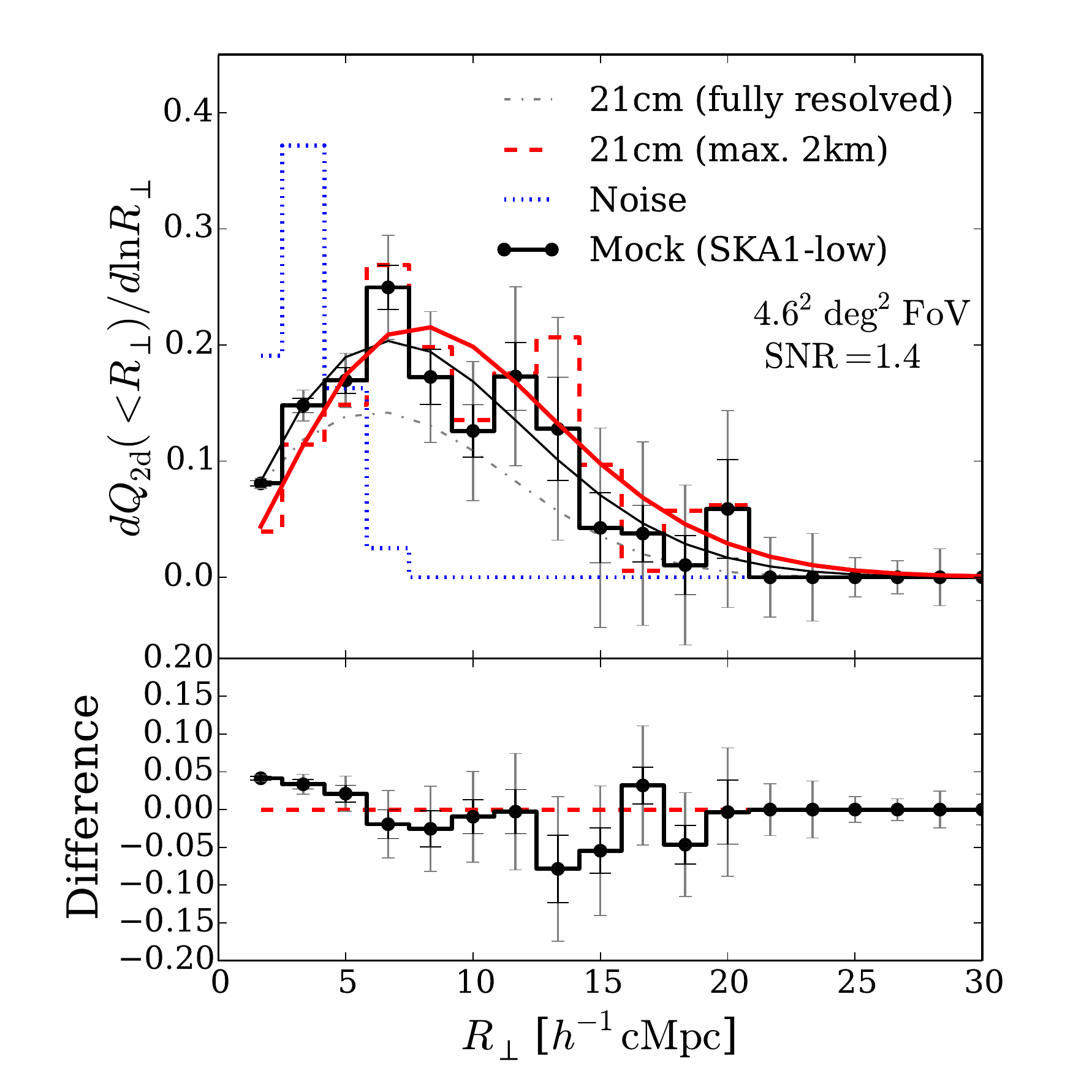}
  \includegraphics[angle=0,width=\columnwidth]{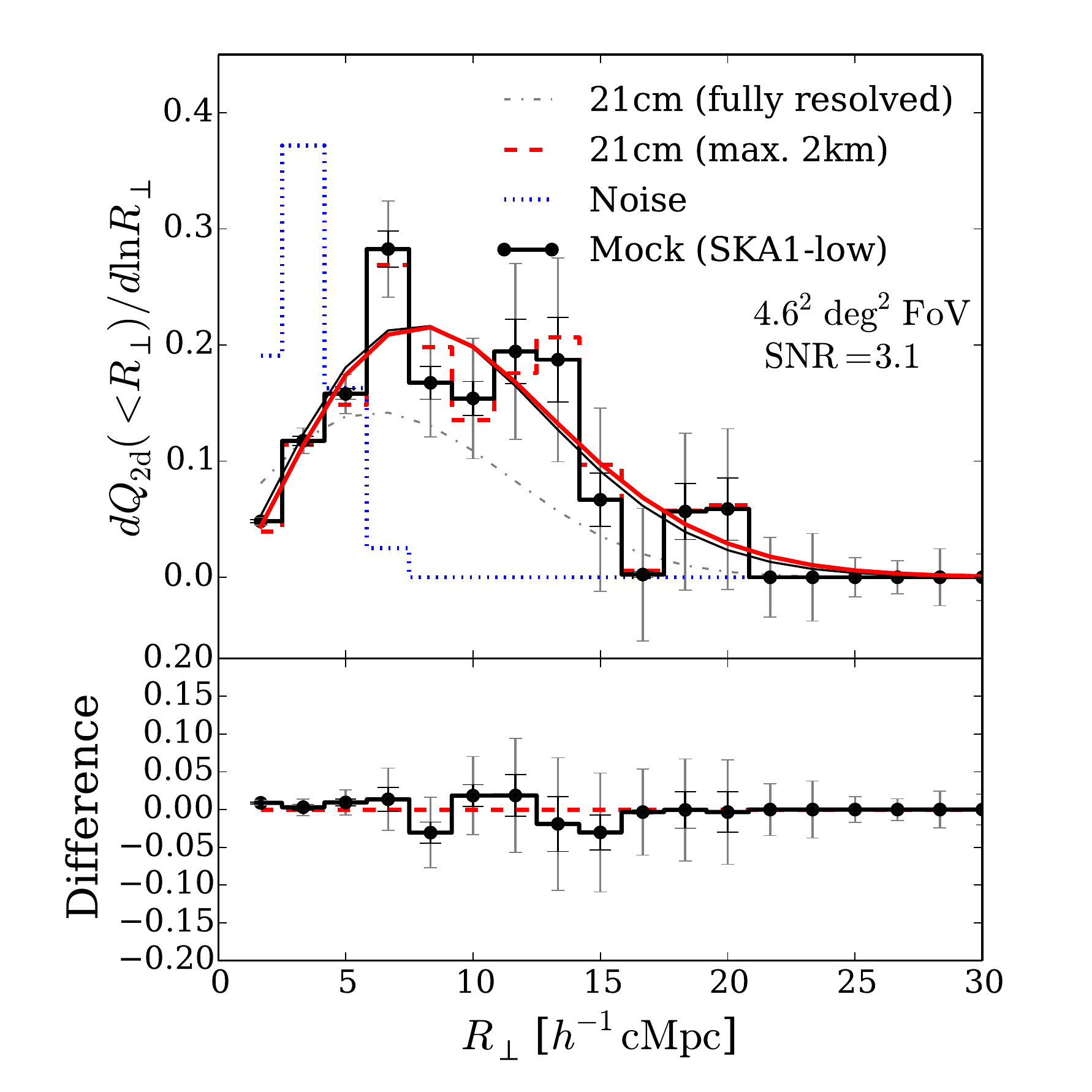}
  \caption{{\it Top:} 21-cm cold-spot size distributions measured from a noiseless 2D image with a 2 km baseline angularresolution (red dashed), noise image (blue dotted), and a mock SKA1-low imaging observation with $\sigma_{\rm N}=4.35\rm~mK$ (left panel) and $\sigma_{\rm N}=2.0\rm~mK$ (right panel) rms noise level. The curves show the best-fit modified Schechter functions. {\it Bottom:} absolute difference between the noiseless 21-cm signal and mock observations. The inner error bars indicate the $1\sigma$ statistical uncertainties due to the thermal noise. The outer error bars include $1\sigma$ statistical error from the sample variance. The grey dotted-dash curve is the best-fit modified Schechter function in the fully resolved 21-cm signal.} 
\label{fig:SKA1_2D}
 \end{center}
\end{figure*}

\begin{figure}
  \includegraphics[angle=0,width=\columnwidth]{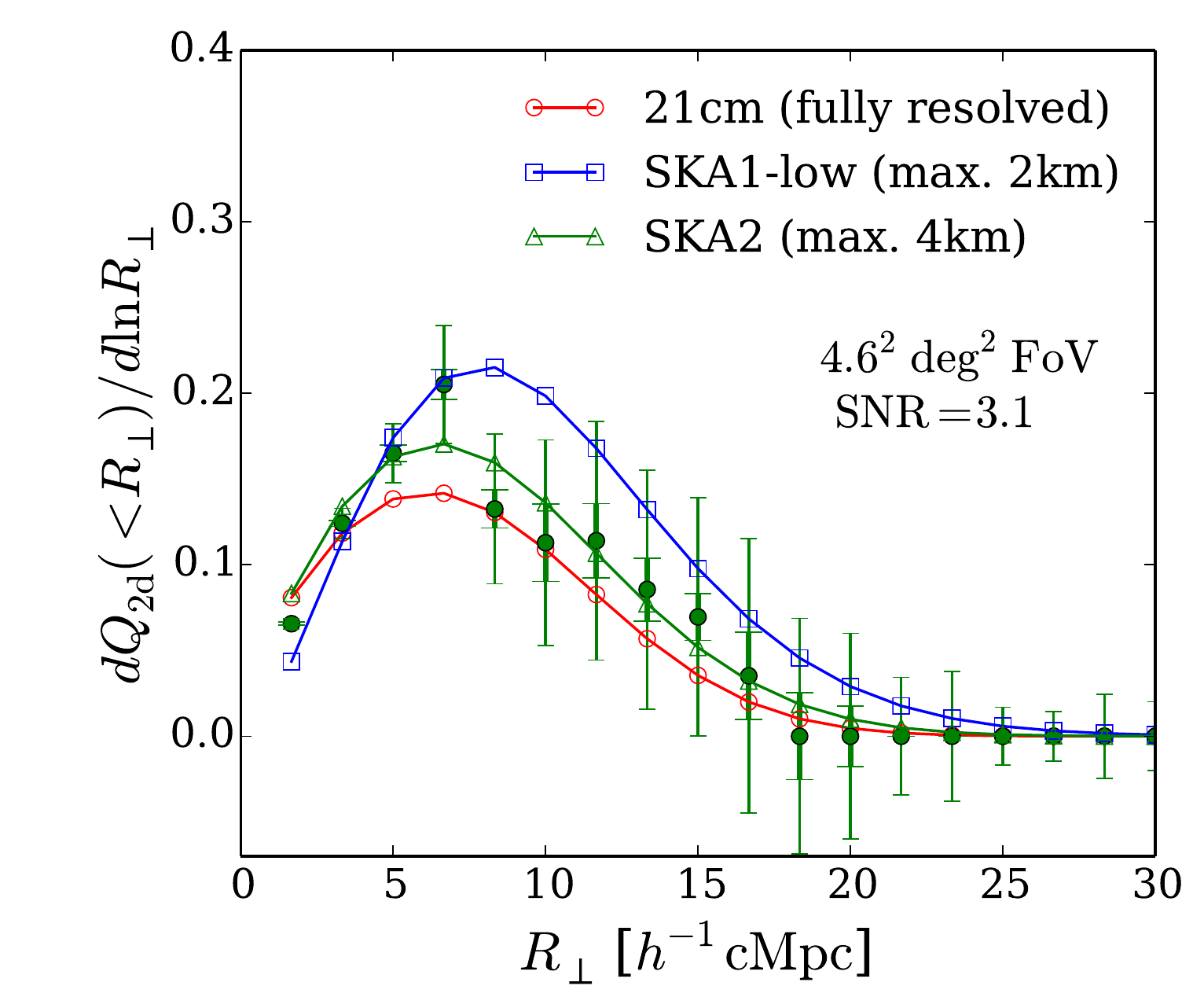}
  \caption{Performance of the graulometric measurement of 21-cm colds spot size distribution once all the requirements are satisfied (see text for details). The green filled points show the size distribution measured from SKA2 with  a $\sigma_{\rm N}=2.0\rm~mK$ noise level. The frequency channel is chosen to be 0.23~MHz wide to match the 1.4 arcmin angular resolution of SKA2. The red, blue, and green curves show the best-fit modified Schechter functions of the size distributions measured from the fully resolved noiseless 21-cm signal, SKA1-low, and SKA2. The inner error bars are only with $1\sigma$ noise error and the outer error bar is the  $1\sigma$ noise+sample variance error.}
\label{fig:SKA2_2D}
\end{figure}

\section{Recovery of $\HII$ region size distribution: 2D data sets}\label{sec:result_2d_granulometry}

In the previous section, we have considered a data cube that contains the full 3D information about the 21-cm signal. In real-world observations, such full 3D information representing a fixed redshift may not be available due to the light cone effect and a bad quality or noise contamination in some frequency slices which must be discarded from the final analysis. Therefore, in this section, we consider more restricted observations of 2D images. We examine the prospects and requirements for recovering the $\HII$ region size distribution from a 2D tomographic image. Choosing only one frequency slice may be too drastic. However, we use this case to illustrate the effects of reducing the frequency width of the 21-cm data set.

The qualitative effects of the field-of-view, angular resolution, and noise on 2D images are the same as the case for 3D data cubes, i.e. (1) a finite FoV introduces a statistical error due to sample variance, but no systematic bias; (2) a limited angular resolution introduces a smoothing bias; and (3) thermal noise introduces a splitting bias, but a small statistical error. However, quantitatively the situation differs because the sample variance becomes extremely large for 2D image and dominates the errors. Therefore, in the following subsections we examine in detail the role of sample variance in 2D images.

\subsection{Effect of the FoV for 2D images}\label{sec:coupling}

The error on the 2D size distribution measurement is sample variance dominated. Figure~\ref{fig:FoV_effect_2D} shows the effect of a finite FoV on the 2D size distribution of 21-cm cold spots obtained from 2D images. 2D radii, $R_{\perp}$, of cold spots are defined as the apparent perpendicular lengths of the cold spots' radii in the 2D image.  We analyse a single frequency image which corresponds to a channel width of 0.45~MHz. For a better recovery of the  size distribution (black solid curve), the granulometric analysis of a larger FoV 21-cm image (red solid histogram, $4.6^2\rm~deg^2$ FoV) is preferred. Averaging over many $2.3^2\rm~deg^2$ FoV single pointing images (black dashed curve) works equally well. The bottom panel shows that the sample variance error of $4.6^2\rm~deg^2$ FoV data is smaller than for the $2.3^2\rm~deg^2$ FoV data by a factor of 2, because of the factor of 4 increase in the area probed. Although the statistical error due to sample variance differs only by a factor of 2, the large end ($R_{\perp}\gtrsim10h^{-1}\rm cMpc$) of the size distribution is difficult to determine by a single pointing FoV image. A large variation (due to the difficulty in sampling many large cold spots) could produce a catastrophic error in the measured 2D size distribution (see blue dashed histogram). While such big error can also occur for large FoV data, the probability is much lower. Therefore, we conclude that increasing the area of the sky probed by increasing the FoV by an interferometric mosaicking or multi-beaming (multiple EoR windows) technique will be important when analysing single frequency 21-cm images (see Section~\ref{sec:discussions} for discussion).

Having a large effective FoV is also important for controlling the thermal noise uncertainty. The statistical uncertainty due to the thermal noise is smaller for a large FoV data as the thermal noise error couples with the FoV. This is because the noise randomly modifies the shape and size of 21-cm cold spots, introducing a statistical uncertainty in the measured size distribution (see Section~\ref{sec:noise_effect}). For a fixed rms noise level, a larger FoV allows the random modifications to be averaged out because there are many cold spots. As the number of larger cold spots is smaller in a small FoV data, the noise error increases for larger cold spot sizes.

Therefore, having a large sky coverage by mosaicking/multi-beaming is also advantageous for reducing the statistical error bars due to the thermal noise. Note that, however, a large sky coverage does not reduce the splitting bias (Section~\ref{sec:noise_effect}), which is only reduced by having a lower absolute value of the rms noise level after a longer integration time.

\subsection{Performance of a successful recovery}

Finally, we test the performance of the granulometric measurement of the 21-cm cold-spot size distribution once all the requirements are satisfied. Figure~\ref{fig:SKA1_2D} shows the expected results for SKA1-low about the granulometric measurement of the 21-cm cold spots using a mock $4.6^2\rm~deg^2$ FoV image at $z=6.8$ after interferometric mosaic imaging by patching many single pointing data. The image is formed with a $0.45\rm~MHz$ channel width. The black points show the mock measurements and the black curve shows the best-fit modified Schechter function. The red curve and histogram refers to an ideal case in the absence of noise. The blue histogram shows the spurious noise cold-spot size distribution.

With a rms noise $\sigma_{\rm N}=4.35\rm~mK$ ($\rm SNR=1.4$ at the resolution element), as shown in the left panel, the splitting bias due to thermal noise is still noticeable on the measured size distribution (black). The splitting bias is within the statistical uncertainties when both sample variance and noise errors are included. For the SKA1-low data with $\sigma_{\rm N}=2.0\rm~mK$\footnote{Although this integration time could unlikely be achieved with SKA1-low for a single FoV, this rms level could be possible with SKA2 phase with a reasonable integration time. One could also apply larger angular or frequency smoothing to enhance the SNR (but must beware of the smoothing bias in the measurement).}, the splitting bias becomes negligibly small. Thus, successfully recovering the 21-cm cold-spot size distribution at 2 km baseline resolution level is possible with the SKA1-low when the mosaicking/multi-beaming technique is used to increase the effective sky coverage.

Figure~\ref{fig:SKA2_2D} shows the expected improvement for SKA2 if it is extended to longer intermediate-scale baselines. The green points are the mock SKA2 measurements, and the green curve with triangles indicates the best-fit modified Schechter function. As already discussed in Section~\ref{sec:mixing_bias}, we should note that the finite angular resolution introduces a fundamental instrumental limitation for SKA1-low (the best-fit modified Schechter function is shown as the blue curve with squares). The smoothing bias systematically shifts the measured size distribution to sizes larger than in the case of a fully resolved 21-cm signal (red curve with open circles). The bias could be as large as a factor of $\sim 2$ for SKA1-low, which exceeds the total statistical uncertainties. For SKA2, assuming the complete uv coverage extends to 4 km baselines, the data with $\sigma_{\rm N}\approx2\rm~mK$ noise at 1.4 arcmin resolution reduce the smoothing bias. At $R_{\perp}>6h^{-1}\rm cMpc$, i.e. well above the angular resolution smoothing scale (${\rm FWHM}\approx2h^{-1}\rm cMpc$), the SKA2 will be able to recover the underlying true cold-spot size distribution (red curve) well within the statistical uncertainties using a $4.6^2\rm~deg^2$ FoV and $\rm SNR\approx3$ image. At this quality, the sample variance error is the dominant source of uncertainty. Thus, increasing sky coverage is necessary to measure the size distribution better. For the smaller end of the size distribution, increasing angular resolution is necessary to prevent a large smoothing bias.

\section{Discussions}\label{sec:discussions}

In this section, we discuss the implications and the importance of measuring the $\HII$ region size distribution from 21-cm tomography to understand the EoR in a wider context. A possible observing strategy and baseline design are also discussed.

\subsection{Synergy between 21-cm power spectra and tomography}\label{sec:synergy}

\begin{figure}
 \centering
  \includegraphics[angle=0,width=\columnwidth]{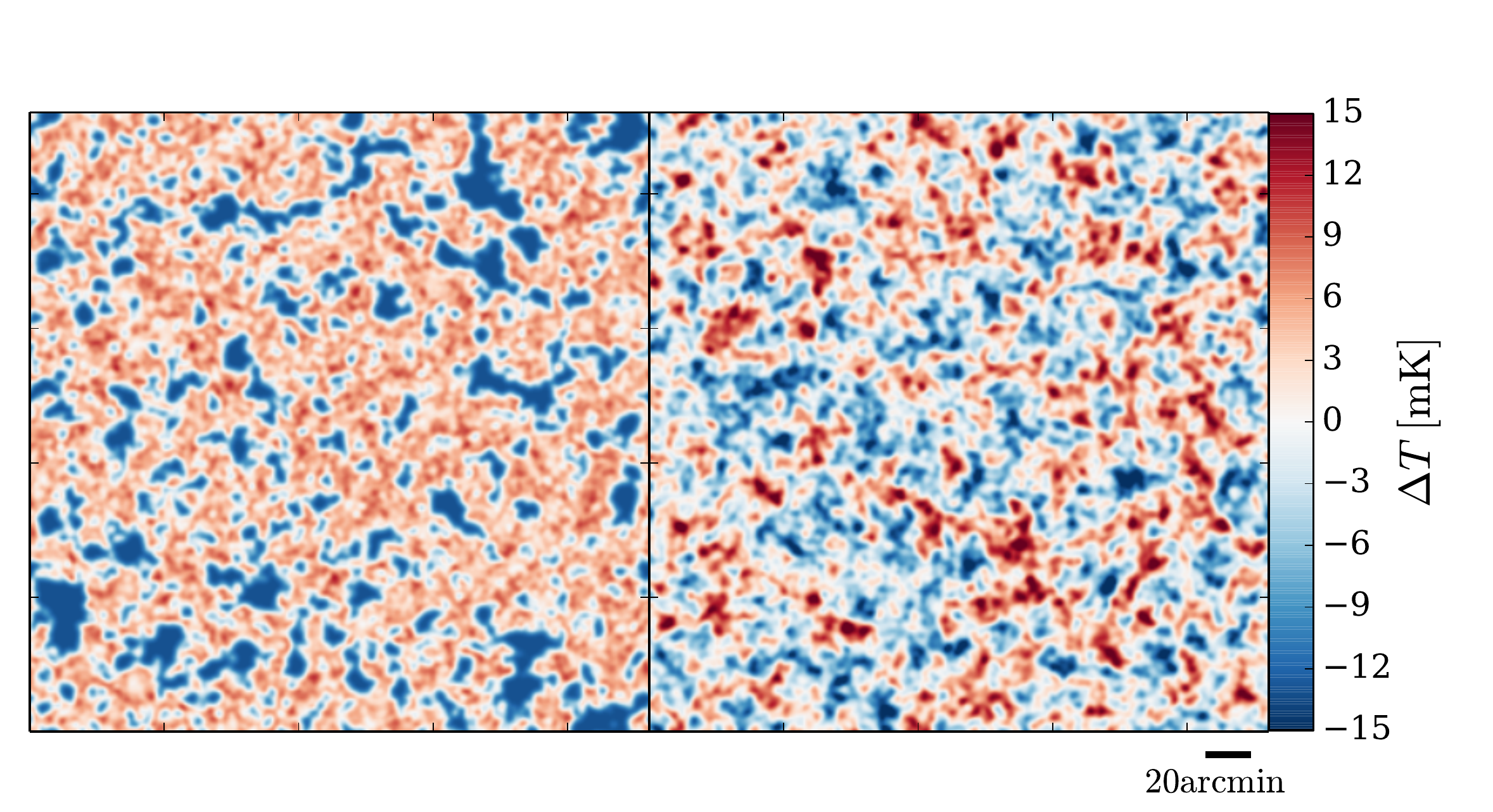}
  \includegraphics[angle=0,width=\columnwidth]{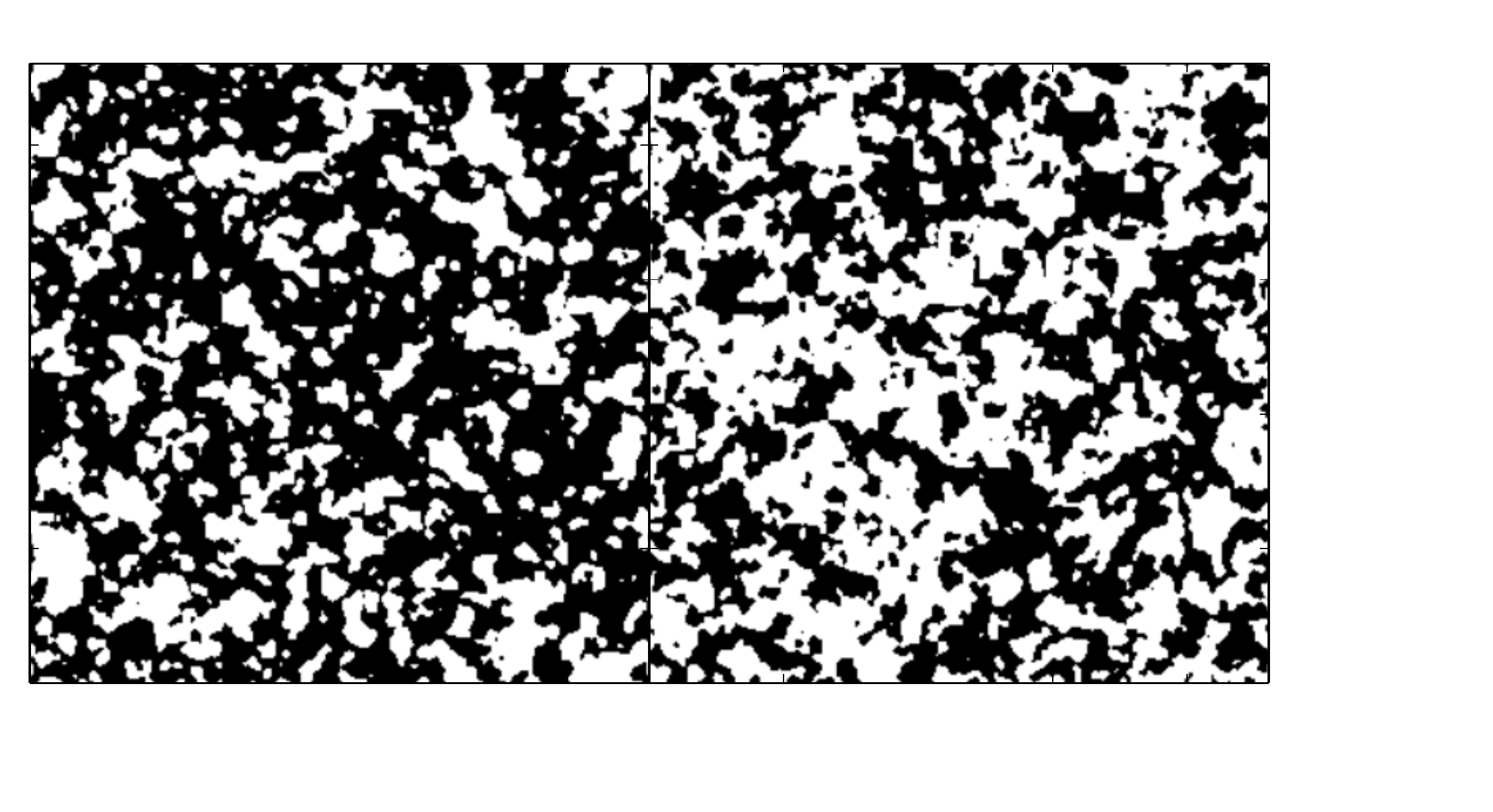}
  \vspace{-1cm}
  \caption{{\it Top}: 2D images of 21-cm signal in the tomography from the RT simulation (left) and a Gaussian random field (right) with the same power spectrum. The mock images are by SKA1-low in the noiseless limit. {\it Bottom}: Binary images of cold spots. The white regions represents the cold spots $\Delta T_{21}<0$.}\label{fig:synergy_map}
\end{figure}

\citet{2015aska.confE..10M} and \citet{2015aska.confE...1K} speculated about the synergy between a 21-cm power spectrum analysis and tomography. Here we present an explicit example to support their arguments.

Figure~\ref{fig:synergy_map} shows two noiseless 21-cm tomographic images with SKA1-low using the RT simulation and a Gaussian random field which has a 21-cm power spectrum identical to the one from the RT simulation. By eye we can discern clear morphological differences between the two images. However, given that their power spectra are identical, how can we tell whether they are two different realizations of the same reionization process or intrinsically different models?

\begin{figure}
 \centering
  \includegraphics[angle=0,width=\columnwidth]{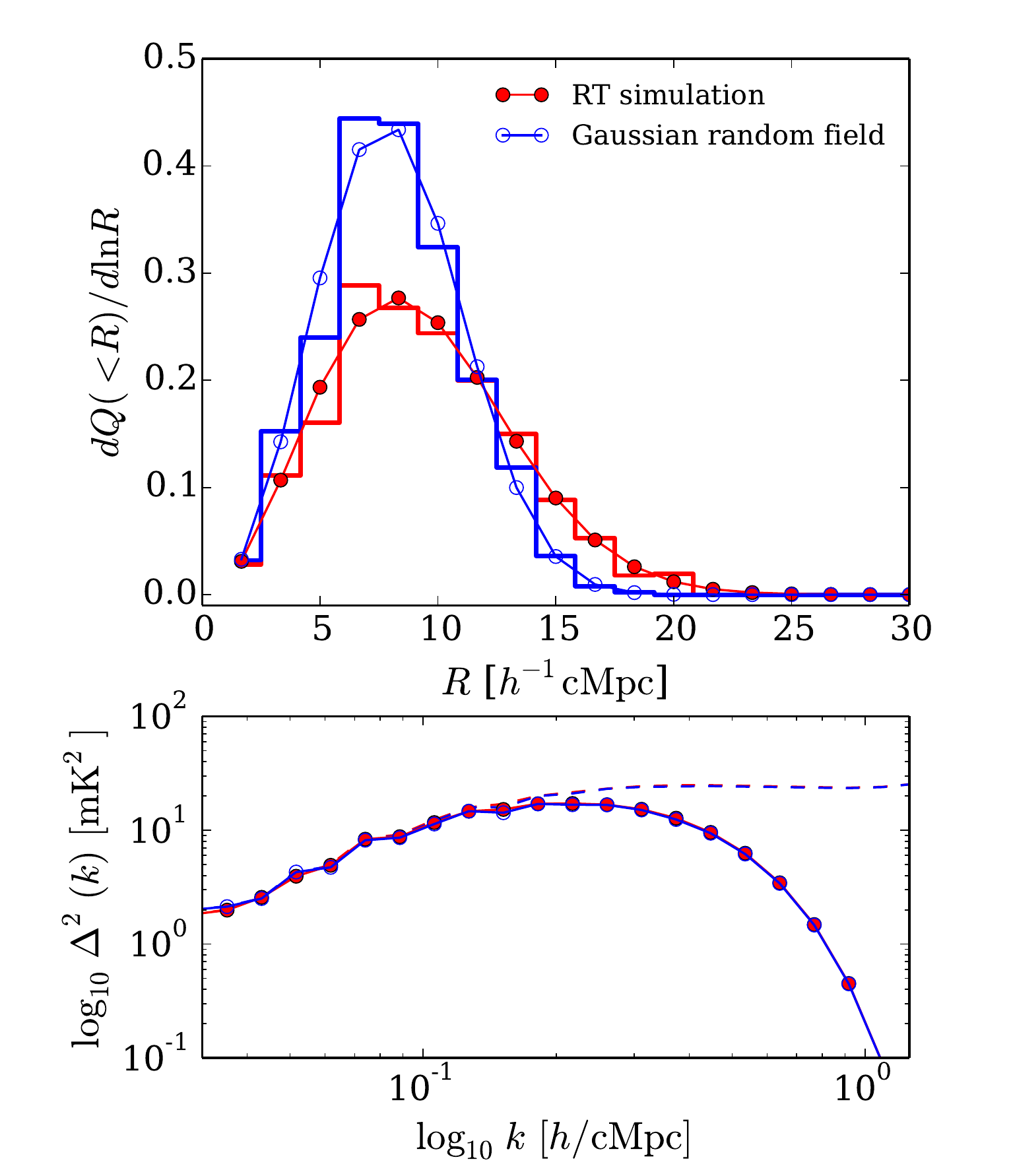}
  \caption{Comparison of cold-spot size distributions for two models with identical power spectra. The results are from the RT simulation (red filled circles) and the Gaussian random field (blue open circles). The histograms are the measured size distributions and the curves are the best-fit modified Schechter functions. The solid curves in the bottom panel show the 21-cm power spectra of the two models measured from the mock SKA1-low in the noiseless limit. The dashed curve shows the case without the effect of angular resolution. The figure clearly illustrates the ability to break the degeneracy in the power spectrum analysis by size distribution measured from 21-cm tomography.}\label{fig:synergy}
\end{figure}

Figure \ref{fig:synergy} shows that the 21-cm cold-spot size distribution measurement from 21-cm tomography using the granulometric analysis has the capability to break this degeneracy in 21-cm power spectrum measurements. This clearly illustrates, although for an extreme example, that tomographic observations add valuable information to the measurement of the 21-cm signal.

Of course, the information from the power spectrum and tomography are complementary to each other. For example, the 21-cm power spectrum contains information about the clustering of $\HII$ regions, which is not captured by size distribution. Furthermore, the higher order multipoles of the power spectrum can potentially quantify the strength and nature of correlation between the underlying matter distribution and the $\HII$ region distribution \citep{majumdar16a,majumdar16}. Therefore, we expect a clear synergy between radio interferometric observations of 21-cm power spectra and imaging tomography. Hence, we place an emphasis on optimizing and balancing the design of future 21-cm experiments for both power spectrum and tomography to achieve an optimal scientific return. 

\subsubsection{Role of intermediate baselines}

\begin{figure}
 \centering
  \includegraphics[angle=0,width=\columnwidth]{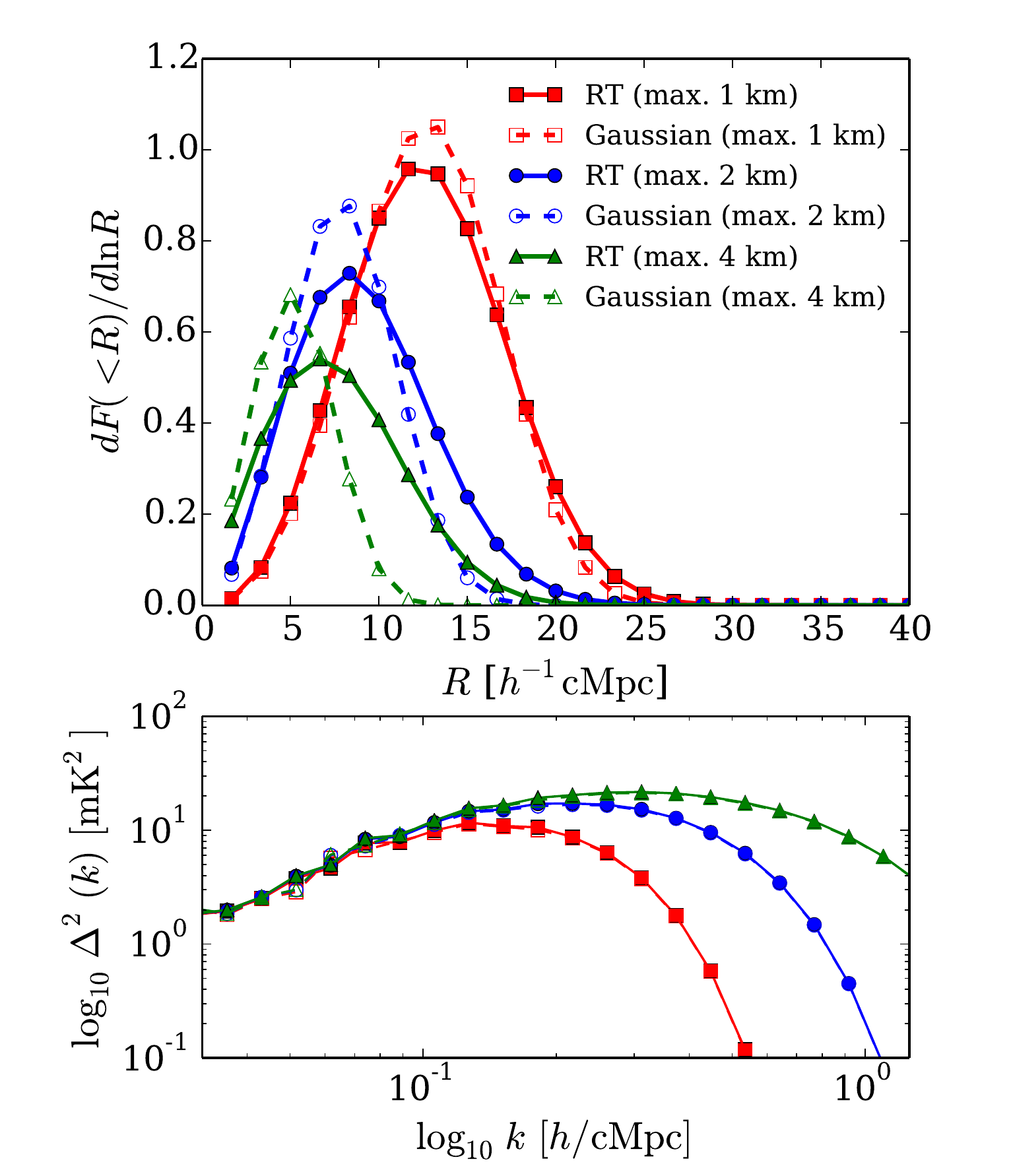}
  \caption{Same as Figure~\ref{fig:synergy},  but varying the maximum baseline used for the mock observation: 1 km (red lines and symbols) and 2 km (blue) for SKA1-low, and 4 km (green) for SKA2. Note that y-axis is shown in $dF(<R)/d\ln R$ to highlight the change in the shape. The figure illustrates the need for longer intermediate baselines for distinguishing the observed 21-cm cold-spot size distribution.}\label{fig:synergy_long}
\end{figure}

How well tomographic data can distinguish different models will of course depend on the (angular) resolution of the images, which is set by the length of the longest baselines used in the construction of the tomographic data. Figure~\ref{fig:synergy_long} shows the effect of angular resolution on the 21-cm power spectrum and the cold-spot size distribution measured from 21-cm tomography. It shows the size distributions and power spectra measured with resolutions $\approx 5.70,~2.85,~1.43\rm~arcmin$. These correspond to maximum baseline lengths of 1 km (red), 2 km (blue), and 4 km (green) at $z=6.8$. Note that all angular resolutions used are smaller than or comparable to the characteristic size [$\sim10h^{-1}\rm cMpc$ ($\sim5.5\rm~arcmin$)] of $\HII$ regions. 

For a mock observation using a maximum baseline of 1~km, the size distributions of both the RT simulation and the Gaussian random field start to resemble each other, as they to coverge to the same shape with decreasing angular resolution. 
In fact, the only difference is due to the volume-filling factor of cold spots. A low angular resolution (large smoothing scale) dictates the shape of the size distribution $dF(<R)/dR$ [and $dQ(<R)/dR$]. The signature of underlying $\HII$ region size distribution is erased, or not significant. While the particular level of angular resolution required to distinguish models differs depending on the stage of reionization, high angular resolution is clearly necessary for differentiating the shapes of size distributions. The power spectra are obviously indistinguishable by design at all angular resolution. 

Therefore, the merit of intermediate baselines and high angular resolution for this particular dataset is clear from Figure~\ref{fig:synergy_long}. Here the measurement of the $\HII$ region size distribution can only break the degeneracy in the 21-cm power spectrum measurements when baselines longer than 2 km are used. Of course this conclusion depends on the actual size distribution of ionized regions. For later stages, where the size distribution peaks at larger sizes, maximum baselines of 1~km would still suffice.

\subsection{Observing strategy and baseline design}

Overall, we suggest that a gradual roll-out of longer intermediate baselines in the SKA2 phase will be beneficial to quantify the reionization morphology from 21-cm tomography. Since the granulometric size distribution measurement can work with a moderate SNR ($\sim3$) imaging, it is more important to sample a large range of $k$-modes in order to eliminate the systematic biases than to increase the SNR of a limited range of large-scale $k$-modes (which may be preferred for power spectra analysis). In this regard, a tiered radio survey \citep{2015aska.confE...1K} combining deep narrow field (a single beam pointing) and shallow wide field (multi-beaming) could also benefit the measurement of $\HII$ region size distributions from 21-cm tomography. The tiered observation would deliver a layer of images with a sufficient SNR and FoV for a given angular scale for the $\HII$ size distribution measurement. Therefore, using the deep-narrow and shallow-wide field images, we may be able to separately measure the small and large ends of the size distribution, which could make effective use of the integeration time and compromise with other science cases such as the power spectrum measurement. In addition, we expect a sweet spot in the baseline configuration, as the splitting bias from noise and the smoothing bias from finite angular resolution have an opposite effect on the size distribution. Further investigation of optimal observing strategy and baseline design for 21-cm tomography will certainly benefit the SKA EoR science cases. Naturally, the first aim should be a detection of a direct 21-cm image at large scales with compact dense core stations, and a large-scale pilot imaging is certainly valuable for testing the imaging algorithm/calibration/foreground removal for 21-cm tomography. Once the pilot phase is over, a larger gain will be achieved by extending to longer intermediate baselines, which would provide a robust physical interpretation and the unbiased size distribution of $\HII$ regions.

\subsection{Comparison with other size measures}\label{sec:comparison}

The literature contains several other algorithms for characterizing the size distribution of $\HII$ regions during reionization, e.g. (1) volumes of topologically connected regions, or the equivalent radii of these (a.k.a. the Friends-of-friends method) (\citealt{2006MNRAS.369.1625I,2008ApJ...681..756S}); (2) the largest radius at which the average ionized fraction around a pixel is above a certain threshold (e.g.\ 90\%) (spherical average method) (\citealt{2007ApJ...654...12Z,2007MNRAS.377.1043M}); (3) distance from a random pixel in $\HII$ regions to the edge of the $\HII$ region containing the pixel (mean free path method) (\citealt{2007ApJ...669..663M}); (4) equivalent radius of Watershed basin (Watershed algorithm) (\citealt{2016MNRAS.461.3361L}).

In the terminology introduced by \cite{2016MNRAS.461.3361L}, the mean free path and spherical average methods are both biased and diffusive in that even in the case of an ensemble of equally sized, non-overlapping, spherical $\HII$ regions, they produce a range of sizes \citep{2016MNRAS.461.3361L}. The other two methods, just as granulometry, would reproduce the correct, unbiased size distribution.
 
The Friends-of-friends method focuses on the volume of connected regions, which means that due to the percolative nature of reionization, already at around 10\% ionization most of the ionized volume will be part of one large region. It therefore is more suitable to test the percolation process \citep{2016MNRAS.457.1813F}. 

The watershed method was recently proposed by \citet{2016MNRAS.461.3361L}. It is commonly used in image processing and it segments the data into discrete bubbles. \citet{2016MNRAS.461.3361L} showed it to produce good bubble size distributions. However, unlike granulometry, it requires an additional tunable parameter to suppress Poisson noise.

The reason that so many different methods have been proposed is because there is no unique way to capture with a simple size distribution the complex morphology of ionized regions during reionization. \cite{friedrich11} and \citet{2016MNRAS.461.3361L} compared these different methods and explored the similarities and differences between their results. \citet{2013ApJ...767...68M} showed that the matched filter technique can be used to measure the bubble size distribution although the associated systematic bias is still to be understood. Each of these methods can in principle be used to derive its answer for the size distribution in real data and use this as a metric to compare to simulation results. However, because they each use a different approach we should not compare the results between different methods. Future studies will reveal which method(s) are most useful to constrain reionization parameters from 21-cm tomographic data.

\section{Conclusions}\label{sec:conclusions}

Using 21-cm tomography one can directly measure a fundamental quantity of the reionization process, i.e. the size distribution of the $\HII$ regions, in a model-independent way. The central questions that we have dealt with in this paper are -- {\it What can we learn from 21-cm tomography using a next generation radio interferometer such as the SKA?} {\it What are the observational requirements for achieving a good science return from such a 21-cm tomography?} We summarize our main conclusions as follows.

\begin{itemize}
\item Granulometric analysis of the 21-cm tomographic data allows the measurement of the $\HII$ region size distribution.
\end{itemize}

We have introduced a novel technique, called ``granulometry'', to quantify the $\HII$ region size distribution in a mathematically well formulated framework, but which is in practice rather simple to implement. The technique attempts to trace the underlying probability distribution function of the $\HII$ region sizes. This places the previously not so well defined concept of ``the morphology of reionization'' on a firm mathematical foundation. Using mock interferometric observations of RT simulations, we have shown that the granulometric analysis of 21-cm tomographic data allows us to recover the $\HII$ region size distribution with the SKA; the theoretical and observational systematics, requirements, and observing strategies have also been examined in detail.

The measured size distribution is well described by a modified Schechter function. Even with our simplest application of the granulometric analysis, the 21-cm cold spots work as an excellent tracer of $\HII$ regions during the second half of reionization (i.e. $Q_{\HII} \geq 0.4$). The size statistics of cold spots can therefore be directly interpreted as that of the $\HII$ regions. For the first half of reionization, attention must be payed to the possible confusion between $\HII$ regions and the voids of density fluctuations in the 21-cm cold-spot size distribution. Although this is not a fundamental limitation of the granulometric analysis, an improved method to separate them for a physical interpretation of the $\HII$ region size statistics must be used in future work.

\begin{itemize}
\item Observational requirements to recover the $\HII$ region size distribution from 21-cm tomography are attainable with the SKA. An ideal observing strategy would be a moderate signal-to-noise, wide-field mosaic/multi-beamed imaging with additional longer intermediate baselines ($\sim2-4\rm~km$).
\end{itemize}

To measure the cold-spot size distribution using 21-cm tomography of an actual radio interferometric observation, we considered two limiting cases: (1) the measurement from a 3D image cube and (2) the measurement from a 2D image slice. The fundamental requirements for recovering the true cold-spot size distribution is to have a high angular resolution (i.e. intermediate baselines of $\sim2-4\rm~km$) and low sample variance (i.e. large sky coverage), both of which are achievable by the SKA by employing an appropriate observational strategy. Each of the requirements and systematics are discussed below.

A finite angular resolution, even if it is below the characteristic scale of $\HII$ regions, introduces the most important source of systematic bias (`smoothing bias') for the measured 21-cm cold-spot size distribution {\it at all scales}. This smoothing bias skews the true size distribution to larger sizes. Although the current baseline distribution of SKA1-low (maximum 2 km baseline) limits the bias below approximately $\sim50~$\%, a higher angular resolution is desirable for an accurate unbiased measurement. Controlling the smoothing bias is a key to successfully recover the statistical characterization of sizes and morphology of $\HII$ regions.

The sample variance due to a finite field-of-view and frequency sampling is a main source of statistical uncertainties. Analysis of a 3D image cube suffers little from the finite field-of-view because of the large frequency samples. The data from a single-pointing observation of SKA1-low allows to measure the cold-spot size distribution with a small statistical uncertainty. On the other hand, for the analysis of a 2D image slice from 21-cm tomography, the sample variance must be reduced to small enough values for a reasonable measurement.  A single-pointing observation suffers from a large sample variance in this case. In fact, because of the propagation of sample variance to the statistical uncertainty of the thermal noise, lowering the sample variance is also advantageous to reduce the statistical error. Mosaicking/multi-beaming techniques for 21-cm imaging tomography will be desirable to increase the sky coverage (i.e. to lower sample variance). 

The thermal noise is not a major obstacle. This can be already manageable with SKA1-low. A moderate SNR ($\lesssim3$) of 21-cm images still permits the recovery the cold-spot size distribution within $\sim50\%$. The required rms noise level is $\sim4\rm~mK$, which should be achievable by the SKA1-low data. 
The largest observational systematics is the bias, instead of a statistical error, caused by the thermal noise (`splitting bias'). The splitting bias artificially skews the measured size distribution to smaller sizes. However, its effect becomes negligible even for $\rm SNR\approx 3$ imaging (with the rms noise level of $\sim2~\rm mK$). A long, but manageable, integration time 
reduces the effect of noise to a negligible level.

\begin{itemize}
\item Requirements for the science beyond power spectra and the synergy between 21-cm power spectrum and tomography. 
\end{itemize}

Our results put a significant importance on having intermediate baselines in radio interferometers. This important factor should be taken into account for the design of a future extension of SKA1-low to SKA2, and other radio telescopes. The intermediate baselines are needed to fully exploit the synergy between 21-cm power spectrum and tomography. For science beyond power spectra, the availability of intermediate baselines is a pre-requisite for distinguishing the size distributions and morphology of $\HII$ regions across different reionization models (Section~\ref{sec:synergy}). 

The intermediate baselines are necessary for reducing a fundamental instrumental  limitation by the angular smoothing bias. In addition, they may be beneficial for overcoming a more immediate challenge of calibration such as the scintillation noise due to ionospheric turbulence. Provided that the future extension of SKA will invest on additional longer intermediate baselines,  a promising observing strategy for 21-cm tomography is an interferometric mosaicking/multi-beaming imaging. 

Finally, the granulometric analysis introduced in this paper is only the tip of the iceberg of the entire spectrum of tools in mathematical morphology and stochastic geometry. They provide powerful means for quantifying the morphology of reionization based on a firm mathematical foundation and theory. Armed with this foundation, the synergy between 21-cm power spectrum and tomography provides many opportunities for directly probing the reionization morphology and extending 21-cm science beyond power spectrum.

\section*{Acknowledgments}

KK acknowledge support from Richard Ellis and the European Research Council Advanced Grant FP7/669253. KK would also like to thank for the hospitality offered by the Department of Astronomy at Stockholm university during one visit, where this work was initiated. Some of the work was also conducted during the workshop supported by the Munich Institute for Astro- and Particle Physics (MIAPP) of the DFG cluster of excellence ``Origin and Structure of the Universe''.
SM would like to acknowledge financial assistance from the European Research
Council under ERC grant number 638743-FIRSTDAWN. We acknowledge
PRACE for awarding us computational time under PRACE4LOFAR
grants 2012061089 and 2014102339 and access to resource Curie
based in France at CEA and to resource SuperMUC at LRZ. This work was supported by the Science and Technology Facilities Council [grant numbers ST/F002858/1 and ST/I000976/1]; and The Southeast Physics Network (SEPNet).

\bibliographystyle{mnras}
\bibliography{Reference}

\appendix
\section{Basic operations in mathematical morphology}\label{sec:App}

Mathematical morphology \citep{Matheron,Serra1983} defines a set of operations to formulate an {\it algebra of shapes}. In this appendix we present an intuitive explanation of these basic morphological operations using diagrams. For a mathematical rigorous treatment, the reader is referred to \citet{Matheron}, \citet{Serra1983}, \citet{Dougherty} and \citet{Chiu2013}.

Suppose that a binary field (image), $X$, is probed by a symmetric structuring element, $S$. There are four elemental operations as follows.

\begin{figure}
\centering
  \includegraphics[angle=0,width=0.98\columnwidth]{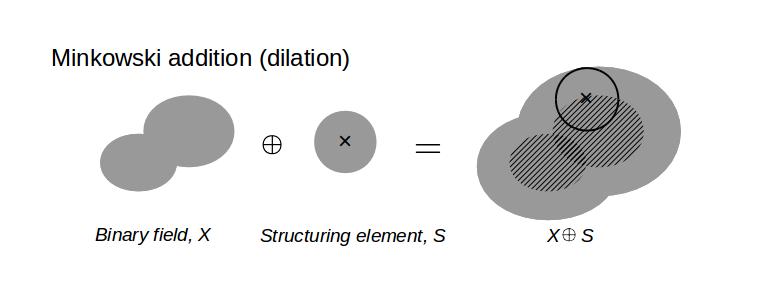}
    \includegraphics[angle=0,width=0.98\columnwidth]{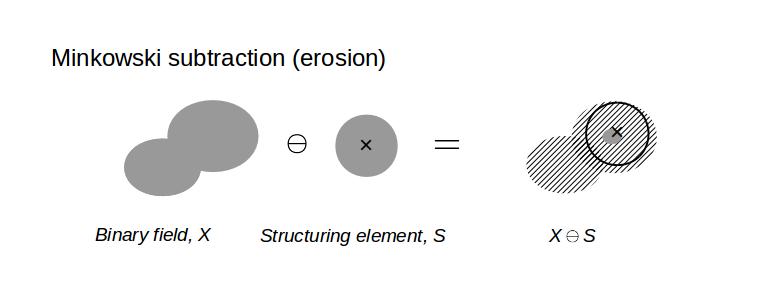}
      \vspace*{-0.4cm}
  \caption{Diagrams describing how the Minkowski addition and subtraction work. The gray areas indicate the resulting binary images. The centre of the structuring element $S$ is indicated by a cross. The hatched region and the circle are drawn as a guide.}\label{fig:Minkowski}
\end{figure}

The Minkowski addition (dilation)\footnote{Strictly speaking, dilation (erosion) and Minkowski addition (subtraction) are not an identical operation. The two operations are identical when a structuring element is symmetrical. Since we employ a sphere as a structuring element in this paper, we use the two terms interchangeably.}, denoted by a symbol $\oplus$, is defined as the union of a binary field, $X$, and a structuring element, $S$, as the centre of the structuring element is moved inside the binary field. Figure~\ref{fig:Minkowski} (top) shows how this operation works. The Minkowski addition enlarges the orignal binary field with a structuring element. 

The Minkowski subtraction (erosion), denoted by a symbol $\ominus$, is a dual to the Minkowski addition. It is defined as the intersection of a binary field, $X$, with the centre of a structuring element, $S$, as the structuring element moves inside the binary field. Figure~\ref{fig:Minkowski} (bottom) shows how the Minkowski subtraction works. Only the parts that can fit the structuring element inside the binary field remain. The Minkowski subtraction shrinks the original binary field with a structuring element.

\begin{figure}
\centering
  \includegraphics[angle=0,width=0.95\columnwidth]{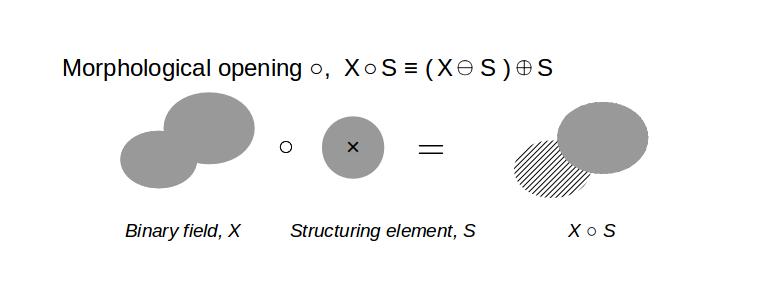}
  \includegraphics[angle=0,width=\columnwidth]{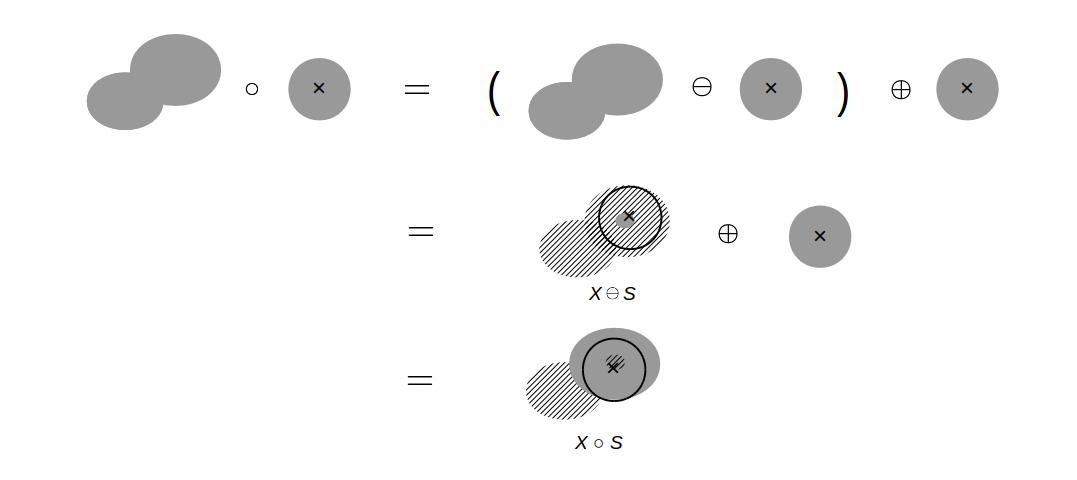}
      \vspace*{-0.5cm}
\caption{Diagrams describing how the morphological opening works. The meaning of the colour and symbols are same as in Figure~\ref{fig:Minkowski}.}\label{fig:opening}
\end{figure}

The morphological opening, denoted by a symbol $\circ$, is then defined by a consecutive operation: Minkowski subtraction followed by Minkowski addition. The morphological opening of a binary field, $X$, by a structuring element, $S$, is expressed as $X\circ S\equiv (X\ominus S)\oplus S$. Figure~\ref{fig:opening} (top) shows the initial and final results of the morphological opening operation. A step-by-step algebra of shapes defined in this framework of mathematical morphology is shown in Figure~\ref{fig:opening} (bottom).

The morphological closing is a dual operation to the morphological opening. Although here we do not use this operation, we define it for completeness. The morphological closing, denoted by a symbol $\bullet$, is defined by a consecutive operation: Minkowski addition followed by Minkowski subtraction, $X\bullet S\equiv (X\oplus S)\ominus S$. 

\vspace*{-0.9cm}

\label{lastpage}

\end{document}